\documentclass[useAMS,usenatbib]{mn2e}

\usepackage{lscape}
\usepackage{aas_macros}
\usepackage[pdftex]{graphicx}

\newbox\grsign \setbox\grsign=\hbox{$>$} \newdimen\grdimen \grdimen=\ht\grsign
\newbox\simlessbox \newbox\simgreatbox \newbox\simpropbox \newbox\wtildebox 
\setbox\simgreatbox=\hbox{\raise.5ex\hbox{$>$}\llap
    {\lower.5ex\hbox{$\sim$}}}\ht1=\grdimen\dp1=0pt
\setbox\simlessbox=\hbox{\raise.5ex\hbox{$<$}\llap
    {\lower.5ex\hbox{$\sim$}}}\ht2=\grdimen\dp2=0pt

\newcommand{\be}{\mbox{\begin{equation}}}
\newcommand{\ee}{\mbox{\end{equation}}}

\newcommand{\Cref}{\mbox{$m_{\rm ref}$}}

\newcommand{\msun}{\mbox{M$_\odot$}}

\renewcommand{\d}{{\rm d}} %roman d in equations

\title{Modeling the formation and evolution of star cluster populations in galaxy simulations}   
\author{J.~M.~Diederik Kruijssen,$^{1,2,\star}$ F.~Inti~Pelupessy,$^2$ Henny~J.~G.~L.~M.~Lamers,$^1$ \newauthor Simon~F.~Portegies~Zwart$^2$ and Vincent~Icke$^2$\\
$^{1}$Astronomical Institute, Utrecht University, PO Box 80000, 3508 TA Utrecht, The Netherlands; $^\star${\tt kruijssen@astro.uu.nl}\\
$^{2}$Leiden Observatory, Leiden University, PO Box 9513, 2300 RA Leiden, The Netherlands}

\begin{document}

\date{Accepted 2011 February 3. Received  2011 February 2; in original form 2010 November 11.}

\pagerange{\pageref{firstpage}--\pageref{lastpage}} \pubyear{2010}
\label{firstpage}

\maketitle

\begin{abstract}
The formation and evolution of star cluster populations are related to the galactic environment. Cluster formation is governed by processes acting on galactic scales, and star cluster disruption is driven by the tidal field. In this paper, we present a self-consistent model for the formation and evolution of star cluster populations, for which we combine an $N$-body/SPH galaxy evolution code with semi-analytic models for star cluster evolution. The model includes star formation, feedback, stellar evolution, and star cluster disruption by two-body relaxation and tidal shocks. The model is validated by a comparison to $N$-body simulations of dissolving star clusters. We apply the model by simulating a suite of 9 isolated disc galaxies and 24 galaxy mergers. The evolutionary histories of individual clusters in these simulations are discussed to illustrate how the environment of clusters changes in time and space. It is found that the variability of the disruption rate with time and space affects the properties of star cluster populations. In isolated disc galaxies, the mean age of the clusters increases with galactocentric radius. The combined effect of clusters escaping their dense formation sites (`cluster migration') and the preferential disruption of clusters residing in dense environments (`natural selection') implies that the mean disruption rate of the population decreases with cluster age. This affects the slope of the cluster age distribution, which becomes a function of the star formation rate density (star formation rate per unit volume). The evolutionary histories of clusters in a galaxy merger vary widely and determine which clusters survive the merger. Clusters that escape into the stellar halo experience low disruption rates, while clusters orbiting near the starburst region of a merger are disrupted on short timescales due to the high gas density. This impacts the age distributions and the locations of the surviving clusters at all times during a merger. The paper includes a discussion of potential improvements for the model and a brief exploration of possible applications. We conclude that accounting for the interplay between the formation, disruption, and orbital histories of clusters enables a more sophisticated interpretation of observed properties of cluster populations, thereby extending the role of cluster populations as tracers of galaxy evolution.
\end{abstract}

\begin{keywords}
galaxies: evolution -- galaxies: interactions -- galaxies: star clusters -- galaxies: starburst -- galaxies: kinematics and dynamics -- galaxies: stellar content
\end{keywords}

\section{Introduction} \label{sec:intro}
It is one of the central aims in current astrophysics to constrain the formation and evolution of galaxies, and their assembly through hierarchical merging \cite[e.g.][]{sanders96,kennicutt98,cole00,conselice03,kauffmann03,vandokkum05,mcconnachie09,hopkins10}. Galaxy mergers play a fundamental role in hierarchical cosmology \citep{white78}, introducing the evolution of the galaxy population as a prime tool to verify cosmological models \citep[e.g.][]{kauffmann93,somerville99,bell05}. Since the late 1980s, observational studies have uncovered a wealth of stellar clusters in galaxy interactions. Because star clusters are easily observed up to distances of several tens of megaparsecs, it is often said that star clusters can be used to probe the formation and evolution of galaxies. This would enable the reconstruction of the merger histories of their parent galaxies \citep{schweizer87,ashman92,schweizer98,larsen01,bastian05}.

The differences between populations of young (massive) star clusters and globular cluster systems show the impact of nearly a Hubble time of evolution \citep[e.g.][]{elmegreen97,vesperini98,vesperini01,fall01,kruijssen09b}, under the assumption that globular clusters initially shared most of the properties of current young star clusters \citep[e.g.][]{elmegreen97}. These differences suggest that cluster populations can indeed be used to trace galaxy evolution, especially because their formation and evolution are known to be governed by their galactic environment \citep{spitzer87,ashman92,baumgardt03,lamers05a,gieles06}. It is therefore crucial to assess {\it how} a cluster population is affected by environmental effects.

There have been substantial efforts in theoretical studies to describe the formation and evolution of star cluster systems. Possible formation sites of star clusters in general and globular clusters in particular have been addressed in theoretical studies \citep[e.g.][]{harris94,elmegreen97,shapiro10} and numerical simulations \citep{bekki02,li04,bournaud08,renaud08}. These studies all point to gas-rich environments with high pressures and densities as the possible formation sites of rich star cluster systems. However, they do not reproduce populations of star clusters and globular clusters that are presently observed, because they focus on cluster formation and either contain only a very simplified description for star cluster evolution or none at all. As they age, star clusters leave their primordial regions and dynamically decouple from the gas of their formation sites. More importantly, star clusters experience extensive dynamical evolution after their formation, which shapes the characteristics of the star cluster populations that are observed today.

Theoretical and numerical studies on the evolution of star clusters have shown that clusters dissolve due to two-body relaxation in a steady tidal field \citep[e.g.][]{spitzer87,fukushige00,portegieszwart01,baumgardt03} and due to heating by tidal shocks \citep[e.g.][]{spitzer58,ostriker72,chernoff86,spitzer87,aguilar88,chernoff90,kundic95,gnedin97,gieles06}. This dynamical evolution leaves a pronounced imprint on the population that survives disruption. In particular, the age and mass distributions of star cluster populations have emerged as excellent tools to trace the disruption histories of clusters \citep[e.g.][]{vesperini01,fall01,lamers05,prieto08}. This implies that the strength of the disruption processes will determine how and to what extent the characteristics of evolved cluster populations still trace the conditions of their formation.

The census of the formation and evolution of star clusters has been applied to populations of star clusters in several studies that focus on the modeling of the observed cluster age and mass (or luminosity) distributions \citep[e.g.][]{elmegreen97,boutloukos03,hunter03,gieles05a,lamers05,smith07,larsen09}. These studies show that the disruption rate of star clusters varies among different galaxies, and also that peaks in the age distributions of star clusters can be used to trace interaction-induced starbursts. Interestingly, the galaxy sample for which the typical disruption rates have been derived shows higher disruption rate for galaxies with high star formation rates.

The above analyses of the formation and disruption of cluster populations are based on two key assumptions:
\begin{itemize}
\item[(1)] The formation rate of stars and clusters is assumed to be constant throughout a galaxy and often also in time. If not assumed constant in time, it is parameterised with a simple function, often a sequence of step functions.
\item[(2)] The disruption rate of star clusters is assumed to be constant throughout a galaxy and in time.
\end{itemize}
While these assumptions can be made for a first-order approach to the formation and evolution of star cluster populations, it is known from theoretical principles of star formation and cluster disruption that they do not hold in actual galaxies. Particularly the observation that the disruption rate of star clusters varies among different galaxies shows that it should also vary within a galaxy: for individual clusters as they pass through different galactic regions, but also for the entire cluster population as a galaxy evolves. The variation with time and space of the cluster formation and disruption rates may or may not affect the observable properties of star cluster populations.

Galaxy interactions provide a clear example of the formation and destruction of cluster populations and the dependence thereof on the local galactic environment. Large numbers of star clusters are formed in the nuclear starbursts occurring during galaxy interactions \citep{holtzman92,whitmore95,schweizer96,whitmore99}. These starbursts are triggered by the angular momentum loss of the gas in the progenitor galaxy discs and the subsequent inflow of the gas \citep{hernquist89,mihos96,hopkins06}. As a result, the gas density in certain locations within galaxy mergers is very high. It was shown by \citet{gieles06} that star clusters are efficiently disrupted by the tidal shocks that arise when they gravitationally interact with passing giant molecular clouds (GMCs). Because the GMC density in central regions of galaxy mergers is high, we should expect tidal disruption to be important. This leaves us with the possible, intriguing combination of the enhanced formation and destruction of star clusters during certain episodes in the hierarchical buildup of galaxies \citep[see][]{kruijssen10}. The violent circumstances in the nuclear region contrast with the outer regions of a galaxy merger, where cluster formation and destruction proceed more temperately.

In general, the galactic conditions under which star cluster populations have been formed and have evolved over the history of the universe may have varied widely \citep[e.g.][]{reddy09}. The use of star cluster populations to trace galactic histories would therefore benefit from a thorough understanding of the co-formation and co-evolution of galaxies and star clusters. Such a census can only be achieved by simultaneously assessing all relevant physical mechanisms, i.e. combining the formation and evolution of star clusters in a single model and allowing for variations with environment.

A thorough way of modeling cluster evolution would be to perform collisional $N$-body simulations of the evolving cluster population in a changing galactic tidal field. This has been done in several studies, for a limited range of cluster masses, orbits and tidal histories \citep[e.g.][]{vesperini97b,baumgardt03,praagman10}. However, these papers only consider the effects of smooth potentials and ignore the disruptive effect of GMCs and other substructure in the gas distribution, thereby limiting the application of such models to globular cluster systems and extremely gas-poor galaxies. Moreover, $N$-body modeling is computationally so expensive that it is not possible to calculate the evolution of the millions of clusters that are formed during the lifetime of a galaxy. In order to self-consistently compute the formation and evolution of an entire cluster population, the best approach would be to implement semi-analytic cluster models in numerical simulations of galaxies. $N$-body simulations of dissolving clusters in time-dependent tidal fields can then be used to provide benchmarks for the semi-analytic cluster evolution models.

In this paper, we aim to self-consistently model the formation and evolution of cluster populations in numerical simulations of (interacting) galaxies. For that purpose, we have combined semi-analytic star cluster models \citep[{\tt SPACE},][]{kruijssen08,kruijssen09c} with an $N$-body/Smoothed Particle Hydrodynamics (SPH) code for modeling galaxies \citep[{\tt Stars},][]{pelupessy04,pelupessy05}. As discussed above, the physical mechanisms that play a role in the formation and evolution of star clusters have all been studied separately in detail. Combining them should allow us to obtain a better picture of how different galactic environments affect their star cluster populations. 

\citet{prieto08} combined cosmological simulations with a semi-analytic description for the dynamical evolution of globular clusters. Their aim was to model the population of globular clusters from high redshift to the present day, and not the self-consistent modeling of the the entire star cluster population including a range of destruction and formation mechanisms in different galactic environments. Particular examples of differences with our approach are the following:
\begin{itemize}
\item[(1)] We include clusters with masses down to a fiducial lower mass limit of 100~\msun. In \citet{prieto08}, only clusters with initial masses $M_{\rm i}>10^5~\msun$ are considered. While this does not influence the present day globular cluster system due to the destruction of lower mass clusters over nearly a Hubble time of evolution, it does obstruct the modeling of young (globular) cluster populations, in which the cluster masses do go down to a physical lower mass limit of around 100~\msun \citep[see e.g.][]{portegieszwart10}.
\item[(2)] In our simulations, the clusters are continuously formed at the sites of star formation that are calculated in the $N$-body/SPH simulation. This is one of the main aspects of our model that enables a self-consistent modeling of the formation and evolution of the cluster population. Conversely, \citet{prieto08} assume that the initial distribution of clusters follows the distribution of the baryonic surface density, which does not necessarily match sites of star and cluster formation.
\item[(3)] In our cluster evolution model, the dynamical mass loss rate of clusters due to two-body relaxation depends on environmental effects, because it is known that the tidal field strength determines the mass loss rate \citep[see e.g.][]{baumgardt03}. This aspect of cluster disruption was not included by \citet{prieto08}.
\end{itemize}
Smaller differences include the exact prescription for mass loss from clusters due to tidal shocks and the evolution of the cluster half-mass radius.

The paper is structured as follows. In Sect.~\ref{sec:model}, we first discuss the simulation code, divided in a section on the galaxy ($N$-body/SPH) model, and a section on the derivation, construction and testing of the star cluster evolution model. The model runs are summarised in Sect.~\ref{sec:runs}, while some key results are presented in Sects.~\ref{sec:discs} (isolated disc galaxies) and~\ref{sec:mergers} (galaxy mergers). The paper is concluded with a summary and a discussion of possible improvements and potential applications of the model.

Throughout the paper, we adopt a standard $\Lambda$CDM cosmology and follow the consensus values after the WMAP results \citep[e.g.][]{spergel07}, with vacuum energy and matter densities $\Omega_\Lambda=0.7$ and $\Omega_{\rm M}=0.3$, and present-day Hubble constant $H_0=70$~km~s$^{-1}$~Mpc$^{-1}$.

\section{Simulation code} \label{sec:model}
Our simulations are performed using a collisionless $N$-body/SPH code in which the formation and evolution of star clusters are included as a sub-grid model. The simulated galaxies contain particles for the stellar, gaseous and dark matter components.

\subsection{Galaxy model}
The evolution of the stellar and dark matter components are governed by pure collisionless Newtonian dynamics, calculated using the Barnes-Hut tree method \citep{barnes86}. The particles sample the underlying phase space distribution of positions and velocities and are smoothed on length scales of approximately 0.2~kpc to maintain the collisionless dynamics and to reduce the noise in the tidal field (which is used for the cluster evolution, see~\ref{sec:clevo}). The Euler equations for the gas dynamics are solved using smoothed particle hydrodynamics, a Galilean invariant Langrangian method for hydrodynamics based on a particle representation of the fluid \citep{monaghan92}, in the conservative formulation of \citet{springel02}. This is supplemented with a model for the thermodynamic evolution of the gas in order to represent the physics of the interstellar medium (ISM). Photo-electric heating of UV radiation from young stars is included (assuming optically thin transport of non-ionizing photons). The UV field is calculated from stellar UV luminosities derived from stellar population synthesis models \citep{bruzual03}. Line cooling from eight elements (the main constituents of the ISM H and He as well as the elements C, N, O, Ne, Si and Fe) is included. We calculate ionization equilibrium including cosmic ray ionization. Further details of the ISM model can be found in \citet{pelupessy04} and \citet{pelupessy05}.

Star formation is implemented by using a gravitational instability criterion based on the local Jeans mass $M_{\rm J}$:
\begin{equation}
\label{eq:jeanscrit}
M_{\rm J} \equiv \frac{\pi \rho}{6} \left( \frac{\pi s^2}{G \rho} \right)^{3/2} < M_{\rm ref} ,
\end{equation}
where $\rho$ is the local density, $s$ the local sound speed, $G$ the gravitational constant and $M_{\rm ref}$ a reference mass scale (chosen to correspond to observed giant molecular clouds). This selects cold, dense regions for star formation, which then form stars on a timescale $\tau_{\rm sf}$ that is set to scale with the local free fall time $t_{\rm ff}$: 
\begin{equation}
\label{eq:tausf}
\tau_{\rm sf}=f_{\rm sf} t_{\rm ff}= \frac{f_{\rm sf}}{\sqrt{4 \pi G \rho}} ,
\end{equation}
where the delay factor $f_{\rm sf} \approx 10$. Numerically, the code stochastically spawns stellar particles from gas particles that are unstable according to Eq.~\ref{eq:jeanscrit} with a probability $1-\exp{(-\d t/\tau_{\rm sf})}$. The code also assigns a formation time for use by the stellar evolution library, and sets the initial stellar and cluster population mass distributions (see below). Mechanical heating of the interstellar medium by stellar winds from young stars and supernovae is implemented by means of pressure particles \citep{pelupessy04,pelupessy05}, which ensures the strength of feedback is insensitive to numerical resolution effects. In this way, the global efficiency of star formation is determined by the number of young stars needed to quench star formation by UV and supernova heating, which is set by the cooling properties of the gas and the energy input from the stars.

\subsection{Star cluster model}
\subsubsection{Cluster formation} \label{sec:clform}
Star clusters are formed in the simulations when a Jeans-unstable gas particle produces a star particle as described above. It is presently not possible to simulate clusters as individual particles for the full cluster mass range, because even with modern computational facilities it would require following too many particles to allow reasonable computation times. Therefore, we choose to spawn the star clusters as the ``sub-grid'' content of a new-born star particle. Their masses are generated from a power law distribution with an exponential truncation at high masses \citep{schechter76}:
\begin{equation}
\label{eq:cimf}
N\d M\propto M^{-2}\exp{(-M/M_\star)}\d M ,
\end{equation}
where $N$ is the number of clusters, $M$ is the cluster mass, and $M_\star=2.5\times 10^6~\msun$ is the adopted exponential truncation mass, which is needed to explain the present day mass function of Galactic globular clusters \citep[see e.g.][]{kruijssen09b}. We allocate 90\% of the new-born particle mass for the star clusters (the ``cluster formation efficiency'' or CFE). Because we adopt a single value of the CFE for all particles, its precise value is not important and merely acts as a normalisation of the number of clusters. The remaining 10\% of the mass is considered to be born in unbound associations of stars that immediately disperse into the field after they are formed\footnote{We do not distinguish between the formation of unbound field stars and the early disruption of star clusters during gas expulsion, which is known as ``infant mortality'' \citep{lada03,goodwin06}.}. This dispersion does not occur physically in the simulation, because the star particle contains both the field stars and the star clusters. All stars have masses distributed according to a \citet{kroupa01} IMF in the mass range 0.08~$\msun$--$m_{\rm max}$, where $m_{\rm max}$ is the maximum stellar mass at $\log{(t/{\rm yr})}=6.6$ (which is the minimum age of the adopted stellar evolution models, see Sect.~\ref{sec:clevo}).

Owing to the sub-grid nature of the cluster implementation, the maximum mass of the formed star clusters is determined by the gas particle mass, star formation efficiency\footnote{This is the fraction of the gas mass that is used to form the star particle.} and CFE, as the mass of a single cluster can not exceed the mass of the star particle it is part of. As a result, the typical maximum cluster mass is $\log{(M/\msun)}\approx 5.8$. We have chosen the number of particles in the simulation such as to optimally cover the cluster mass range of interest and to have sufficient numerical resolution. An algorithm that allows for simultaneous star formation in groups of gas particles is being included in a future study.

\subsubsection{Cluster evolution} \label{sec:clevo}
After their formation, star clusters evolve by losing mass by stellar evolution and dynamical evolution. The star cluster evolution is computed with the {\tt SPACE} models \citep{kruijssen08,kruijssen09c}, which include a semi-analytic description of the evolution of cluster mass and its stellar content. They include stellar evolution from the Padova models \citep{marigo08}, stellar remnant production, remnant kick velocities \citep[e.g.][]{lyne94}, dynamical disruption \citep{lamers05} and the evolution of the stellar mass function owing to the stellar mass dependence of the ejection rate of stars from the cluster \citep{kruijssen09c}. The total mass loss rate is constituted by the contribution from stellar evolution, $(\d M/\d t)_{\rm ev}$, and the contribution from tidal disruption, $(\d M/\d t)_{\rm dis}$:
\begin{equation}
\label{eq:dmdt}
\left(\frac{\d M}{\d t}\right) = \left(\frac{\d M}{\d t}\right)_{\rm ev} + \left(\frac{\d M}{\d t}\right)_{\rm dis} .
\end{equation}
The mass loss due to stellar evolution is computed using the Padova isochrones by following the decrease of the maximum stellar mass over one timestep, and integrating the mass function within the cluster over the corresponding mass interval. {This method assumes the instantaneous removal of stars and ignores the gradual nature of stellar winds. Stars typically only lose mass during the last $\sim 10$\% of their lifetime, during which the maximum stellar mass decreases by merely a few percent, and the total cluster mass by even less. The mass loss rates of the most massive stars are also comparable during the enclosed time interval, which implies that the instantaneous removal of the most massive stars is balanced by the delay of mass loss from slightly less massive stars. This ensures that the obtained mass loss rate is very similar to the actual rate due to stellar winds and supernovae at any time \citep[see e.g.][]{kruijssen08}.} We include the production and retention of stellar remnants as discussed in \citet{kruijssen09c}.

The mass loss rate due to disruption is driven by two-body relaxation and tidal shocks:
\begin{equation}
\label{eq:dmdtdis}
\left(\frac{\d M}{\d t}\right)_{\rm dis} = \left(\frac{\d M}{\d t}\right)_{\rm rlx} + \left(\frac{\d M}{\d t}\right)_{\rm sh} ,
\end{equation}
where the subscripts `dis', `rlx', and `sh' denote disruption, two-body relaxation and tidal shocks, respectively. We now describe the contributions from both mass loss mechanisms\footnote{We neglect a third mass loss mechanism, namely the dynamical mass loss that is induced by the shrinking of the Jacobi radius resulting from the mass loss due to stellar evolution \citep{lamers10}. This is allowed if clusters initially do not fill their Roche lobes. In the irregular tidal fields that we are considering, the Jacobi radius constantly changes. This implies that the equilibrium situation of a cluster filling its Roche lobe is unlikely to occur.}.

The mass loss rate due to two-body relaxation is determined by the strength of the tidal field and the cluster mass \citep{baumgardt03,gieles08}. To describe the mass loss, we adopt the semi-analytic approach of \citet{lamers05} that has been extensively tested against $N$-body simulations of dissolving star clusters and observations \citep[e.g.][]{lamers05a,gieles05a,bastian05,lamers06a}. The mass decreases exponentially on a disruption timescale that decreases as the cluster mass decreases $t_{\rm dis}\equiv(\d\ln{M}/\d t)^{-1}$:
\begin{equation}
\label{eq:dmdtrlx}
\left(\frac{\d M}{\d t}\right)_{\rm rlx}=-\frac{M}{t_{\rm dis}}=-\frac{M^{1-\gamma}}{t_{0}} ,
\end{equation}
where the disruption timescale is given by $t_{\rm dis}=t_0 M^\gamma$. Here, the exponent $\gamma=0.6$---0.8 is the mass dependence of the disruption timescale, and increases with the King parameter $W_0$ of the cluster density profile \citep{baumgardt03,lamers10}. The normalisation constant $t_0$ is the dissolution timescale parameter, which sets the rapidity of the disruption and is determined by the tidal field\footnote{Throughout the paper, we do not only use the term `disruption time(scale)', but also the more intuitive `disruption rate', which is related to the inverse of the timescale. While the disruption timescale is specific to the properties of a cluster and depends on its mass and (for tidal shocks) half-mass radius, the term `disruption rate' is used to refer to the general `disruptiveness' of the environment.}. For clusters on circular orbits in a logarithmic potential, $t_0$ has been related to the angular frequency of the orbit, and subsequently to the ambient density $\rho_{\rm amb}$ as $t_0 \propto \rho_{\rm amb}^{-1/2}$ \citep{baumgardt03,lamers05a}. The physical driving force behind cluster disruption is the tidal field. According to Poisson's law, the tidal field strength $T$ is proportional to the ambient density, implying a more fundamental relation:
\begin{equation}
\label{eq:t0}
t_0=t_{0,\odot}(T/T_\odot)^{-1/2} ,
\end{equation}
where $t_{0,\odot}$ is the dissolution timescale in a logarithmic potential at the galactocentric radius of the sun $R_{\rm gc,\odot}$ and $T_\odot\approx 7.01\times 10^{2}~{\rm Gyr}^{-2}$ is the tidal field strength at that location for a circular velocity of 220~km~s$^{-1}$. For $\gamma=0.62$ one obtains $t_{0,\odot}=21.3$~Myr, while for $\gamma=0.70$ we have $t_{0,\odot}=10.7$~Myr \citep{kruijssen09}. We adopt a density profile with King parameter $W_0=5$ for the clusters and consequently $\gamma=0.62$. {This `typical' King parameter is consistent with observations of open clusters \citep{portegieszwart10} and rapidly dissolving globular clusters \citep{mclaughlin05,kruijssen09}. Clusters with lower King parameters ($W_0\sim 3$) are susceptible to rapid disruption due to stellar evolution-induced mass loss \citep{fukushige95,baumgardt03,lamers10}, while high King parameters of $W_0\ga 7$ are typically achieved after core collapse of the most massive systems such as old globular clusters. To illustrate the influence of the concentration on cluster disruption, we also consider the case of $W_0=7$ in the rest of the derivation of the model.}

To determine the tidal field strength, we first evaluate the tidal field tensor
\begin{equation}
\label{eq:tide}
T_{ij} = -\frac{\partial^2\Phi}{\partial x_i\partial x_j} ,
\end{equation}
where $\Phi$ is the gravitational potential and $x_i$ is the $i$-th component of the position vector. In the simulations, the tidal tensor is computed by numerical differentiation of the force field, which is smoothed on scales of 200~pc, thereby minimising the sensitivity of the evolution of the star clusters to discreteness noise. We use 1\% of the smoothing length for the differentiation interval. The tidal tensor has three eigenvectors, which denote the principal axes of the action of the tidal field. The corresponding eigenvalues represent the magnitude of the force gradient along these axes, with negative eigenvalues denoting compressive components of the tidal field, and positive eigenvalues indicating extensive components \citep[e.g.][]{renaud08}. The tidal field strength $T$, i.e. the quantity that sets the tidal boundary of the cluster, is thus equal to the largest eigenvalue of the tidal tensor. If the tidal field is fully compressive, i.e. all eigenvalues of the tidal tensor are negative, we assume $(\d M/\d t)_{\rm rlx}=0$. The eigenvalues are computed with the routines by \citet{kopp08}.

Tidal shocks disrupt star clusters by increasing the energy of the stars that are bound to the cluster. It was shown by \citet{kundic95} that the first- and second-order energy inputs induced by tidal shocks contribute more or less equally to the disruption of star clusters, while higher-order terms can be neglected. For the mass loss rate due to tidal shocks we write
\begin{equation}
\label{eq:dmdtsh}
\left(\frac{\d M}{\d t}\right)_{\rm sh}=-\frac{M}{t_{\rm sh}} ,
\end{equation}
where $t_{\rm sh}$ denotes the disruption time for tidal shocks. It can be separated in the disruption times due to the first- and second-order energy input, $t_{\rm sh,1}$ and $t_{\rm sh,2}$:
\begin{equation}
\label{eq:tsh}
t_{\rm sh} = \left(t_{\rm sh,1}^{-1}+t_{\rm sh,2}^{-1}\right)^{-1} .
\end{equation}
Both components of the disruption time depend on several properties of the cluster and its environment, and will change over time. 

The derivation of $t_{\rm sh}$ has been treated extensively in literature \citep[e.g.][]{spitzer58,spitzer87,ostriker72,kundic95,gnedin97,gieles07,prieto08}, though the details of these approaches differ. For example, some studies correctly observe that not all of the energy input by the tidal shock is converted into mass loss \citep{gieles07}, while others add the second-order disruption component $t_{\rm sh,2}$ \citep{kundic95}. Also, most studies consider the tidal perturbation of clusters on closed orbits, for which the frequency of disc and bulge shocks is predictable. However, for more erratic tidal shocks, $t_{\rm sh}$ should be linked to the tidal field \citep{prieto08}. Especially when modeling the evolution of star clusters in galaxy mergers this is an important improvement. We follow the lines of most of the above studies, and combine their refinements.

We first compute the first-order disruption timescale due to tidal shocks, which can be expressed as \citep[e.g.][]{gieles07}:
\begin{equation}
\label{eq:tsh1}
t_{\rm sh,1}=\frac{\Delta t}{f}\left|\frac{E}{\Delta E}\right| ,
\end{equation}
where $E$ denotes the cluster energy\footnote{This is the sum of the internal kinetic energy and the internal potential.} per unit mass and $\Delta t$ is the time since the previous shock. The parameter $f$ is the fraction of the relative energy change that is converted to a change in cluster mass. This number is smaller than unity, because stars escape the cluster with velocities above the escape velocity. It is defined as $f\equiv \d \ln{M}/\d \ln{E}$, and has been found to be $f\simeq 0.25$ for two-dimensional (2D) shocks\footnote{Most tidal shocks occur in the orbital plane of the interaction between the cluster and the perturbing object, e.g. a GMC. This corresponds to a 2D shock. A 1D shock resembles a head-on encounter with the perturbing object, which is relatively rare compared to a more distant passage.} \citep{gieles06}. The internal energy per unit cluster mass $E$ is given by
\begin{equation}
\label{eq:eint}
E = -\frac{\eta GM}{2r_{\rm h}},
\end{equation}
with $\eta\simeq 0.4$ a proportionality constant \citep[e.g.][]{spitzer87}, $G$ the gravitational constant, and $r_{\rm h}$ the half-mass radius of the cluster.

We combine the approaches of \citet{gieles07} and \citet{prieto08} to express the average energy change $\Delta E$ of the ensemble of stars in the cluster as a function of their average radial position $r$ and the tidal heating parameter $I_{\rm tid}$:
\begin{equation}
\label{eq:de1}
 \Delta E = \frac{1}{2}\overline{(\Delta v)^2} = \frac{1}{6}I_{\rm tid} \overline{r^2} .
\end{equation}
The tidal heating parameter is written as a function of the tidal tensor \citep{gnedin99b,prieto08}:
\begin{equation}
\label{eq:itid}
 I_{\rm tid} = \sum_{i,j}\left(\int T_{ij}\d t\right)^2A_{{\rm w},ij}(x) ,
\end{equation}
in which the integration is performed over the duration of the tidal shock for the particular component of the tidal tensor. The factor $A_{{\rm w},ij}(x)$ represents a parameterised version of the Weinberg adiabatic correction \citep{weinberg94a,weinberg94b,weinberg94c}. It is defined for each component of the tidal tensor and describes the absorption of the energy injection by the adiabatic expansion of the cluster. The adiabatic correction depends on the product of the average angular frequency of the stars within the cluster $\overline{\omega}$ and the timescale $\tau_{ij}$ of the shock for the corresponding component of the tidal tensor \citep{gnedin97,gnedin99}:
\begin{equation}
\label{eq:aw}
 A_{{\rm w},ij} = \left(1+\overline{\omega}^2\tau_{ij}^2\right)^{-3/2} .
\end{equation}
The value of $\overline{\omega}$ is constant when expressed in $N$-body units \citep{heggie86,gieles07}, but when converted back to physical units it becomes:
\begin{equation}
\label{eq:omega}
 \overline{\omega} = C_\omega\sqrt{\frac{8\eta^3 GM}{r_{\rm h}^3}} ,
\end{equation}
where $C_\omega$ denotes a proportionality constant, which for King parameters $W_0=\{5,7\}$ is $C_\omega=\{0.68,0.82\}$ \citep{gieles07}. The timescale of the shock $\tau_{ij}$ is the time interval in which the corresponding component of the tidal tensor drops by 39\%, coinciding with the definition of one standard deviation in a Gaussian distribution.

Substitution of Eqs.~\ref{eq:eint}---\ref{eq:itid} in Eq.~\ref{eq:tsh1} now gives the disruption timescale due to the first-order effects of tidal shocks:
\begin{equation}
\label{eq:tsh1b}
t_{\rm sh,1}=\frac{3\eta}{f}\frac{GM}{r_{\rm h}^3}\frac{r_{\rm h}^2}{\bar{r^2}}I_{\rm tid}^{-1}\Delta t ,
\end{equation}
with the ratio $\overline{r^2}/r_{\rm h}^2=\{2,3.5\}$ for King profile parameters $W_0=\{5,7\}$ \citep{gieles06}. This equation should be complemented with the disruption timescale due the second-order energy input, or ``shock-induced relaxation'' \citep{kundic95}, which is expressed as
\begin{equation}
\label{eq:tsh2}
t_{\rm sh,2}=\frac{\Delta t}{f}\left|\frac{E^2}{(\Delta E)^2}\right| ,
\end{equation}
where $(\Delta E)^2$ denotes the stellar ensemble-averaged mean square energy change. Following \citet{kundic95}, we write
\begin{equation}
\label{eq:de2}
 (\Delta E)^2 = \overline{(v\Delta v)^2} = \frac{1}{5}I_{\rm tid} \overline{\omega^2 r^4} .
\end{equation}
The stellar ensemble-average $\overline{\omega^2 r^4}$ then follows:
\begin{equation}
  \label{eq:omega2r4}
  \overline{\omega^2 r^4}=\overline{GM(r)r} \equiv \zeta GMr_{\rm h} ,
\end{equation}
where $M(r)$ represents the mass within radius $r$, and the constant $\zeta$ is defined as
\begin{equation}
  \label{eq:zeta}
  \zeta\equiv\frac{\overline{M(r)r}}{Mr_{\rm h}} .
\end{equation}
For King profile parameters $W_0=\{5,7\}$ the values are $\zeta=\{0.81,1.03\}$ (M.~Gieles, private communication).

Substitution of Eqs.~\ref{eq:eint} and~\ref{eq:de2} in Eq.~\ref{eq:tsh2} now gives the disruption timescale due to the second-order effects of tidal shocks:
\begin{equation}
\label{eq:tsh2b}
t_{\rm sh,2}=\frac{5\eta^2}{4f\zeta}\frac{GM}{r_{\rm h}^3}I_{\rm tid}^{-1}\Delta t = \frac{5\eta}{12\zeta}\frac{\overline{r^2}}{r_{\rm h}^2}t_{\rm sh,1} .
\end{equation}
Using Eq.~\ref{eq:tsh}, the disruption time due to the combined first- and second order effects of tidal shocks then becomes:
\begin{eqnarray}
\label{eq:tshfull}
t_{\rm sh}&=&\left(t_{\rm sh,1}^{-1}+t_{\rm sh,2}^{-1}\right)^{-1} \nonumber\\
                 &=&\left(1+\frac{12\zeta}{5\eta}\frac{r_{\rm h}^2}{\bar{r^2}}\right)^{-1}t_{\rm sh,1} \equiv C_{\rm sh}t_{\rm sh,1} ,
\end{eqnarray}
where for $W_0=\{5,7\}$ we have $C_{\rm sh}=\{0.29,0.36\}$, indicating that the contribution of the second-order energy input is most important for low-concentration clusters. The mass loss due to shocks is applied upon the completion of a shock in any of the components of the tidal tensor. Numerically, this means that $I_{\rm tid}=0$ unless a shock is completed, i.e. one of the components of $|T_{ij}|$ reaches a minimum that is at most 88\% of the preceding maximum\footnote{In a Gaussian distribution, this contrast coincides with the location of 1$\sigma$.}.

We have thus far not defined any prescription for the half-mass radius $r_{\rm h}$. In semi-analytic models that do not contain any information regarding the structural evolution of star clusters, this is often related to the (initial) mass according to a power law relation:
\begin{equation}
\label{eq:rh}
r_{\rm h}=r_{\rm h,4}\left(\frac{M_{\rm (i)}}{10^4~{\rm M}_\odot}\right)^\delta ,
\end{equation}
where $r_{\rm h,4}$ is the half-mass radius of a $10^4~\msun$ cluster, $M_{\rm (i)}$ represents the (initial) cluster mass, and $\delta$ is the power law index. The disruption timescale due to tidal shocks $t_{\rm sh}$ and the adiabatic correction $A_{\rm w}$ both depend on the half-mass radius of the cluster (see Eqs.~\ref{eq:aw} and~\ref{eq:tsh1b}), implying that the value of $\delta$ influences the mass dependence of $t_{\rm sh}$. It is therefore important to include a reliable prescription for the half-mass radius. We have tested several dependences of $r_{\rm h}$ on the initial and present cluster mass when comparing the models to the $N$-body simulations by \citet{baumgardt03}. The best agreement is found for $\delta=0.225$ and $r_{\rm h,4}=4.35$~pc, and when using the present-day mass (see Sect.~\ref{sec:test1}). These parameters are within an acceptable range of the `mass loss-dominated mode' of the radius evolution reported in \citet{gieles10b}.

It should be emphasised that the mass-radius relation quoted in Eq.~\ref{eq:rh} does not have the same meaning as the mass-radius relation that can be observed for real star cluster populations \citep[e.g.][]{larsen04b}. Instead, it approximates the evolution of the half-mass radius for a single cluster, given a certain initial radius. In a population of star clusters, which is constituted by clusters of a range of ages, initial masses, initial radii, and mass loss histories, the resulting mass-radius relation of the entire population may be very different, as it is set by the collection of states in which these different clusters happen to exist at the time of the observation. When using a power law formulation, both types of mass-radius relation will only be similar if the initial half-mass radius of the clusters is also set by the cluster mass as in Eq.~\ref{eq:rh}. For mathematical simplicity, we do choose to set the initial half-mass radius according to Eq.~\ref{eq:rh} (see Sect.~\ref{sec:test1}), but it is not a requirement. This approximation is supported by clusters in $N$-body simulations, which tend towards a well-defined evolutionary sequence \citep{kuepper08}, suggesting that the initial radius may be erased after a couple of relaxation times. It is currently not known how the half-mass radius evolves in the erratic tidal fields of real galaxies, which contain GMCs and spiral arms. We therefore choose to adopt the `conservative' formulation of Eq.~\ref{eq:rh}. In a future work, we will include a more sophisticated evolution of the half-mass radius.

The mass loss rates due to two-body relaxation and tidal shocks are combined with a model to compute the evolution of the stellar mass function of the dissolving clusters \citep{kruijssen09c}. In most cases, two-body relaxation gives a depletion of the mass function at low masses, because low-mass stars have a higher probability to escape than massive stars \citep{henon69,vesperini97,takahashi00,baumgardt03,kruijssen09c}. As a result, the integrated photometric properties of star clusters evolve due to dynamical disruption \citep[e.g.][]{baumgardt03,kruijssen08b}. By including this, we can use the presented models to trace dynamical information of the simulated galaxies down to the stellar level. The stars that are lost from clusters are added to the field star population of the star particle in which the cluster resides.

The star cluster model is implemented in the galaxy evolution code to operate `on the fly', simultaneously with the galactic evolution, rather than having the cluster evolution calculated {\it a posteriori} as in \citet{prieto08}. While this approach is already beneficial because it potentially allows for a two-way interaction between a galaxy and its cluster population, it also implies that the tidal history of each cluster is only saved for the most recent time steps, which improves the memory efficiency of the simulation and allows us to model the full star cluster population all the way down to our adopted minimum mass of 100~\msun.

\subsubsection{Star cluster model testing: method} \label{sec:test1}
\begin{figure*}
\center\resizebox{17.5cm}{!}{\includegraphics{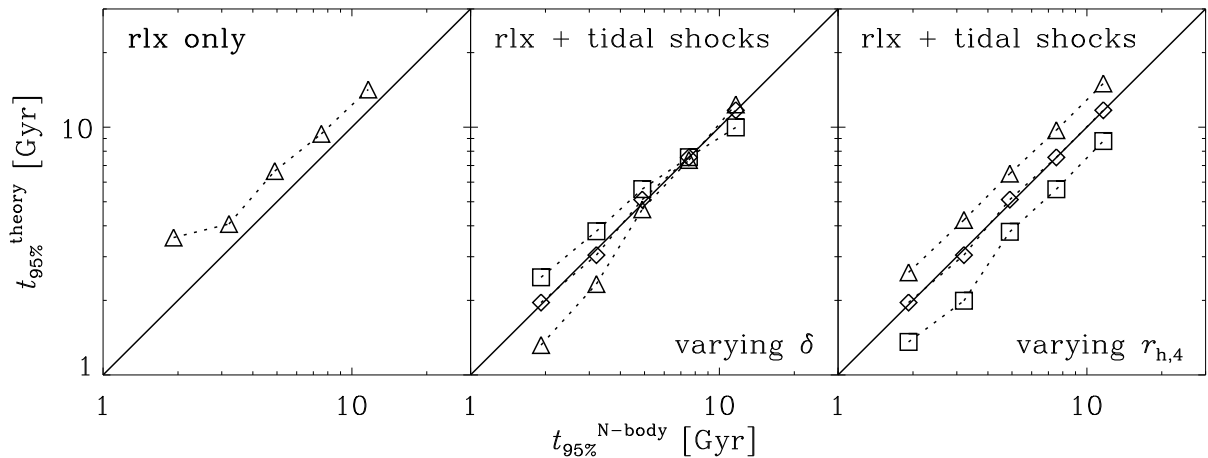}}
\caption[]{\label{fig:t95}
Comparison of our analytically estimated cluster lifetimes with the lifetimes found in $N$-body simulations of clusters on eccentric orbits (eccentricity $e=0.5$) from \citet{baumgardt03}. Connected symbols represent different initial cluster masses, characterised by different numbers of stars (8k, 16k, 32k, 64k, and 128k), while the solid line shows the 1:1 correspondence between our $t_{95\%}$ and that from the $N$-body models. {\it Left}: without tidal shocks. {\it Middle}: Including tidal shocks, for $\delta=\{0.15,{\bf 0.225},0.35\}$ (triangles, diamonds, squares) and $r_{\rm h,4}=4.35$~pc. {\it Right}: Including tidal shocks, for $\delta=0.225$ and $r_{\rm h,4}=\{3,{\bf 4.35},5\}$~pc (triangles, diamonds, squares). The values at which the agreement between both approaches is best are written in boldface and are denoted by diamonds in the figure.
       }
\end{figure*}
We have compared the prescription for star cluster evolution from Sect.~\ref{sec:clevo} to the $N$-body models of dissolving star clusters by \citet{baumgardt03}. These simulations follow the dynamical evolution of initially Roche-lobe filling star clusters in a logarithmic potential with a circular velocity of 220~km~s$^{-1}$. The runs contain clusters on circular and eccentric orbits between galactocentric radii in the range 2.833--15~kpc. The stars in the clusters follow \citet{king66} density profiles with $W_0=5$ or $W_0=7$, and the stellar masses are distributed according to a \citet{kroupa01} initial mass function between 0.1 and 15~\msun.

In this section, we exclusively consider clusters on eccentric orbits, because these clusters experience shocks during each pericentre passage. This allows us to test both disruption mechanisms rather than only two-body relaxation in a steady tidal field, for which the semi-analytic model has been tested extensively in previous studies \citep[e.g.][]{lamers05}. While the mass loss rate due to two-body relaxation contains no free parameters (see Eqs.~\ref{eq:dmdtrlx} and~\ref{eq:t0}), the description for tidal shocks depends on the adopted relation between the half-mass radius and the cluster mass (see Eq.~\ref{eq:rh}), which is governed by the parameters $\delta$ and $r_{\rm h,4}$. Their values are obtained from the comparison.

To compare our models to the simulations from \citet{baumgardt03}, in Fig.~\ref{fig:t95} we show the time after which 95\% of the initial cluster mass is lost ($t_{95\%}$) for our models and for their $N$-body runs. The figure shows poor agreement if only two-body relaxation is included, but good agreement when tidal shocks are accounted for. Additionally, the influence of $\delta$ and $r_{\rm h,4}$ on $t_{95\%}$ is shown. As can be expected from Eq.~\ref{eq:rh}, $\delta$ affects the mass dependence of the disruption time due to shocks (increasing $\delta$ reduces the contrast between the disruption times of different masses), while $r_{\rm h,4}$ impacts the normalisation of the disruption time (compact clusters live longer).

\begin{figure}
\center\resizebox{8cm}{!}{\includegraphics{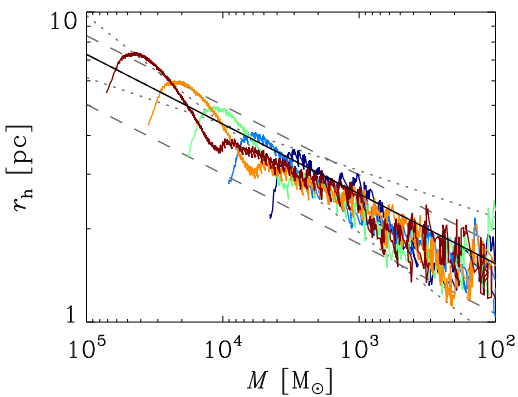}}
\caption[]{\label{fig:mcrh}
Evolution of the half-mass radius as a function of the remaining mass for the $N$-body simulations shown in Fig.~\ref{fig:t95}, with from left to right the initial numbers of stars being 128k, 64k, 32k, 16k, and 8k (coloured irregular lines). The solid line shows our adopted relation between half-mass radius and mass, with $r_{\rm h,4}=4.35$~pc and $\delta=0.225$. The dashed and dotted lines show the variations of $r_{\rm h,4}$ and $\delta$ from Fig.~\ref{fig:t95}, with dashed denoting $r_{\rm h,4}=\{3,5\}$~pc (bottom, top) and dotted denoting $\delta=\{0.15,0.35\}$ (shallow, steep).
       }
\end{figure}

As was mentioned in Sect.~\ref{sec:clevo} and is visible in Fig.~\ref{fig:t95}, the best match between our models and the $N$-body runs is found for $\delta=0.225$ and $r_{\rm h,4}=4.35$~pc. These values should therefore approximate the actual evolution of the half-mass radii in the $N$-body models of the clusters. This is verified in Fig.~\ref{fig:mcrh}, where our adopted mass-radius relation is compared to the actual evolution of the half-mass radii of the clusters in Fig.~\ref{fig:t95}, showing good agreement. The clusters follow evolutionary tracks in the mass-radius plane that are very similar to each other, indicating that the clusters tend to evolve to a common cooling track, analogous to a `main sequence of star clusters' as discussed by \citet{kuepper08}. The obtained mass-radius relation implies that upon losing stars due to disruption, clusters will always slowly evolve towards filling their Roche lobes, because the Jacobi radius depends on mass as $r_{\rm J}\propto M^{1/3}$, implying $r_{\rm h}/r_{\rm J}\propto M^{-0.1}$. The slope of the mass-radius relation is also consistent with the `mass loss-dominated mode' from the work by \citet{gieles10b}. 

In principle, the mass-radius evolution of the clusters could be a relic of the initial conditions of the $N$-body simulations, in which the clusters initially fill their Roche lobes. However, we are considering clusters on eccentric orbits, for which the tidal radius continuously changes, suggesting that whether or not a cluster initially fills its Roche lobe may be irrelevant after a couple of orbits. This would be even more important in more realistic, erratic tidal fields. Most importantly though, the evolution of the half-mass radius shown in Fig.~\ref{fig:mcrh} also includes the time after core collapse, when any possible imprint of the initial conditions will have been erased. Therefore, we do not expect that the details of the initial conditions of the $N$-body simulations would affect the slope or normalisation of the mass-radius relation, especially given its simplicity. {Nonetheless, the mass-radius relation of clusters in erratic tidal fields could deviate from our adopted one. We discuss possible improvements of our approach in Sect.~\ref{sec:improve}.}

\subsubsection{Star cluster model testing: numerical resolution} \label{sec:test2}
For any numerical model, it is necessary to check at which numerical resolutions the results are reliable. Within a realistic galactic environment, the tidal field experienced by star clusters is very erratic, contrary to the well-defined tidal shocks occurring during each pericentre passage in the \citet{baumgardt03} simulations. Testing the resolution requirements of the models (both in time and space) should therefore be done for tidal histories that are taken from our simulations.

\begin{figure}
\center\resizebox{8cm}{!}{\includegraphics{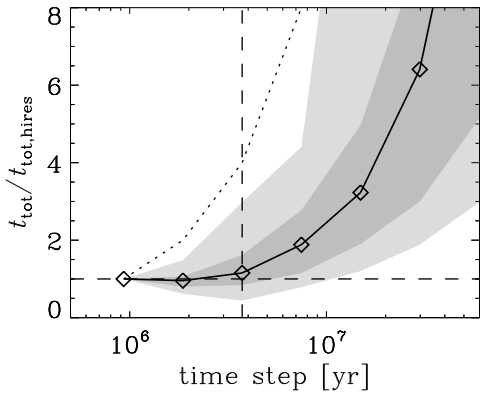}}
\caption[]{\label{fig:tres}
The mean total disruption time of a $2\times 10^4~\msun$ cluster for different time steps, scaled to that for the smallest time step (diamonds/solid line). The horizontal dashed line indicates unity, which coincides with the leftmost diamond per definition. The dark grey tinted area spans the space covered by one standard deviation of the distribution at each time step, while the light grey represents two standard deviations. The dotted line represents the relation for when $t_{\rm tot}/t_{\rm tot,hires}$ were proportional to the time step. The maximum time step used in our simulations is indicated by the vertical dashed line.
       }
\end{figure}
To explore how the time resolution of the tidal field affects our modeled star cluster disruption, we computed the mass evolution of a $2\times10^4~\msun$ cluster for 200 tidal histories that were randomly drawn from the particles in one of our galaxy disc simulations. For each of these histories, the evolution was computed seven times, using fixed time steps of $\{1,2,4,8,16,32,64\}\times 0.932$~Myr. For each time step, we then scaled the total disruption time of the cluster $t_{\rm tot}$ to the total disruption time found for that cluster when using the smallest time step ($t_{\rm tot,hires}$). This ratio can be used to trace the relative change of the total lifetime due to resolution effects. Because tidal shocks are events with a certain duration, some of them could be skipped when increasing the time step, suggesting that in that regime $t_{\rm tot}/t_{\rm tot,hires}$ becomes proportional to the time step.

The mean $t_{\rm tot}/t_{\rm tot,hires}$ of the 200 tidal histories is shown as a function of the time step in Fig.~\ref{fig:tres}. The relation that would be expected if $t_{\rm tot}/t_{\rm tot,hires}$ were proportional to the time step is also included. The figure shows that for large time steps ($\ga 10$~Myr), the total disruption time indeed becomes proportional to the time step, as the durations of some shocks are then short enough to be skipped, while for smaller time steps the total disruption time converges. The maximum time step of the particles in our simulations (3.73~Myr) is such that time resolution effects should not play an important role, particularly because the maximum time step is only used for very weakly accelerated particles in dynamically quiet regions. In the simulations, we do not use fixed time steps, but adaptive ones instead, increasing the resolution as the force on a particle increases \citep{pelupessy04,pelupessy05}, up to a maximum resolution increase of a factor 4096 (potentially yielding a time step of $\sim 1000$~yr). This ensures that tidal shocks, which typically occur when the force on a particle is non-negligible, are always well-resolved. In this way, we minimise the effect of the time step on our computed cluster lifetimes.

Whether or not the evolution of star clusters is affected by the spatial resolution of the simulations depends on the smoothing length and the number of particles used. The distribution of mass needs to be resolved in sufficient detail to ensure that encounters with individual particles do not disrupt the clusters. Such disruption would be artificial, because individual particles are discrete representations of a continuous mass distribution. Whether the resolution requirements are satisfied can be easily checked with an order-of-magnitude estimate.

Most of the disruption due to an encounter with an individual particle would be caused by the corresponding tidal shock. The presented simulations use a smoothing length of $h=200$~pc and typical particle masses of $M_{\rm part}=8\times10^5~\msun$ (see Sect.~\ref{sec:runs}). The typical duration of an encounter with a single particle is then approximately $h/\sigma$, with $\sigma$ the velocity dispersion in a galaxy disc, which is of the order 20~km~s$^{-1}$. This gives a typical shock duration of  about 10~Myr. Since we are interested in an upper limit to the disruptive effect of individual particles, we ignore the adiabatic correction (Eq.~\ref{eq:aw}) and assume that throughout the shock the heating is equal to the tidal heating encountered when the cluster is located at the centre of the particle. For a spline kernel smoothing, the central density of a particle is $\rho_{\rm centre}=M_{\rm part}/(\pi h^3)$. Due to the symmetry of a head-on encounter, the tidal tensor is diagonal with values $T_{ij}=-4GM_{\rm part}\delta_{ij}/(3h^3)$, which for the quoted shock characteristics gives a tidal heating parameter of $I_{\rm tid}\approx10^2~{\rm Gyr}^{-2}$. If this type of shock would be continuously repeated over the entire lifetime of a cluster, it would take well over 120~Gyr to destroy a $10^4~\msun$ cluster. As is evident from Fig.~\ref{fig:t95} and later sections of this paper, such a disruption time is 1--3 orders of magnitude larger than typical disruption times. We conclude that for our choice of particle numbers and smoothing length, encounters with individual particles do not play an important role. Instead, the shocks that lead to the disruption of clusters are produced by groups of particles, such as spiral arms or complexes of molecular clouds, which do have a physical meaning. Consequently, the spatial resolution requirements are satisfied. Note that this strongly depends on the smoothing length $h$, because for the maximum tidal heating we have $I_{\rm tid}\propto T_{ij}^2 \propto M_{\rm part}^2h^{-6}$. This implies that it is not possible to adopt a much smaller smoothing length, which would require require vastly larger numbers of particles to reduce the particle mass and minimise the effect of encounters with individual particles.

\begin{figure}
\center\resizebox{8cm}{!}{\includegraphics{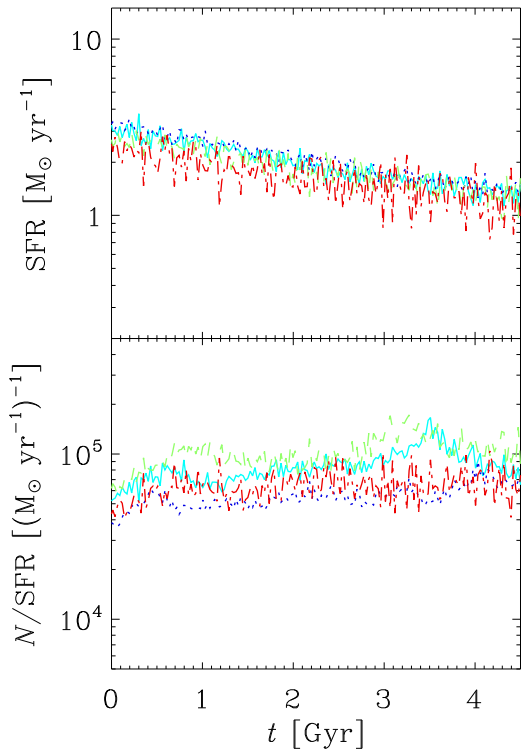}}
\caption[]{\label{fig:spaceres}
Dependence of the simulations on the spatial resolution. {\it Top}: star formation history as a function of time. {\it Bottom}: number of clusters per unit star formation rate as a function of time. Different lines denote simulation 1dB (see Sect.~\ref{sec:runs}) run at different spatial resolutions. Particle numbers of $\{1/4,1/2,1,2\}$ times those used in simulation 1dB are represented by $\{$dash-dotted red, dashed green, solid {cyan}, dotted blue$\}$ lines. The bumps in the bottom panel for simulations 1dB$_{1/2}$ (at $t=3.2$~Gyr), 1dB (at $t=3.5$~Gyr) and 1dB$_2$ (at $t=4.1$~Gyr) occur shortly after the (random) formation of holes in the gas due to feedback effects (see text).
       }
\end{figure}
The stability against resolution effects is illustrated in Fig.~\ref{fig:spaceres}, which shows the dependence of the star formation rate and the number of clusters per unit star formation rate on the spatial resolution. The figure shows that the formation rates of stars and clusters converge with increasing resolution. The bottom panel gives a measure for star cluster disruption, and shows that the variation of the disruption rate with spatial resolution is of the order of the inherent scatter on the number of clusters. The number of clusters very slightly decreases for lower resolutions, because encounters with individual particles then become more important due to their higher masses. Simulations at higher resolutions also exhibit a slight decrease of the number of clusters, because in these the structure in the spatial distribution of the gas is resolved in more detail. While this may induce a small increase of the disruption rate, we do not expect that this continues at even higher resolutions, because the amount of tidal heating scales with the square of the mass of the structure causing the tidal shock, implying that resolving increasingly smaller structures results in a correspondingly smaller addition to the tidal heating\footnote{Even for a population of GMCs that follows a power law mass distribution with index $-2$, the total tidal heating would (linearly) increase with GMC mass, despite the many more low-mass GMCs than massive ones.}. Figure~\ref{fig:spaceres} also illustrates that the mean disruption rate of star clusters is more sensitive to random fluctuations than the overall star formation rate. At certain times, the number of clusters briefly increases due to a decrease of the disruption rate. This is caused by the random, transient excavation of the gas due to feedback in certain star-forming regions, which cause large numbers of clusters to experience less disruption. The mean disruption rate, which depends on the distribution of the gas, then shows larger scatter than the star formation rate, which to good approximation is determined by the mean surface density of the gas.

\section{Summary of the model runs} \label{sec:runs}
\begin{table*}\centering
\caption[]{\label{tab:discs}
     Details of the initial conditions for the disc galaxy models.}
\begin{tabular}{c c c c c c c c c c c c}
\hline
ID & $f_{\rm gas}$ & ${M_{\rm vir}}^a$ & $z$ & $\lambda$ & $N_{\rm halo}$ & $N_{\rm gas}$ & $N_{\rm disc}^{\rm star}$ & $N_{\rm bulge}^{\rm star}$ & ${M_{\rm part}^{\rm halo}}^a$ & ${M_{\rm part}^{\rm bary}}^a$ & Comments
\\\hline
1dA$^b$    &      0.20     &     10$^{12}$     &   2   &     0.05     &     10$^6$     &     10250     &     41000    &      10000   & $10^6$ & $8\times10^5$ &     low gas fraction      \\
1dB$^{b,c,d}$    &      0.30     &     10$^{12}$     &   2   &     0.05     &     10$^6$     &     15375     &     35875    &      10000   & $10^6$ & $8\times10^5$ &  standard model \\
1dC    &      0.50     &     10$^{12}$     &   2   &     0.05     &     10$^6$     &     25625     &     25625    &      10000   & $10^6$ & $8\times10^5$ &     high gas fraction      \\
1dD    &      0.30     &     5$\times10^{11}$ &   2   &       0.05   &       $5\times10^5$   &       7688   &       17938  &        5000 & $10^6$ & $8\times10^5$ &     half mass \\
1dE    &      0.30     &     10$^{12}$     &   2   &     0.05     &     10$^6$     &     15375     &     35875      &  0         & $10^6$ & $8\times10^5$ &     no bulge           \\
1dF    &      0.30     &     10$^{11}$     &   2   &     0.05     &     10$^6$     &     15375     &     35875    &      10000   & $10^5$ & $8\times10^4$ &     low mass           \\
1dG    &      0.30     &     10$^{12}$     &   2   &     0.10     &     10$^6$     &     15375     &     35875    &      10000   & $10^6$ & $8\times10^5$ &     high spin           \\
1dH    &      0.30     &     10$^{12}$     &   0   &     0.05     &     10$^6$     &     15375     &     35875    &      10000   & $10^6$ & $8\times10^5$ &     low concentration        \\
1dI    &      0.30     &     10$^{12}$     &   5   &     0.05     &     10$^6$     &     15375     &     35875    &      10000   & $10^6$ & $8\times10^5$ &     high concentration        \\
\hline
\multicolumn{12}{l}{$^a$In solar masses ($\msun$).}\\
\multicolumn{12}{l}{$^b$To investigate the relative importance of the two disruption mechanisms, these models are also computed for} \\
\multicolumn{12}{l}{disruption excluding tidal shocks (i.e. only two-body relaxation, `1dA/B$_{\rm rlx}$') and for disruption excluding two-body} \\
\multicolumn{12}{l}{relaxation (i.e. only tidal shocks, `1dA/B$_{\rm sh}$').}\\
\multicolumn{12}{l}{$^c$This model is also computed for $i=\{1/4,1/2,2\}$ times the number of baryonic particles (i.e. `1dB$_i$'.)} \\
\multicolumn{12}{l}{$^d$This model is also computed for a constant disruption parameter $t_0=2$~Myr (see Eq.~\ref{eq:dmdtrlx}) and no tidal shocks (`1dB$_{\rm fix}$').}
\end{tabular} 
\end{table*}
We construct model disc galaxies with parameters that can be related to the outcomes of cosmological $\Lambda$CDM galaxy formation models \citep{mo98,springel05b}. They consist of a dark halo with a \citet{hernquist90} profile\footnote{This density profile is very similar to profiles found in cosmological simulations \citep{navarro96,navarro97}. The difference only occurs at radii much larger than the scale radius, where the density profile of the \citet{hernquist90} profile falls of as $\propto r^{-4}$ rather than $r^{-3}$. This does not affect the results in this paper, because our galaxy merger simulations do not include galaxies on very wide orbits.}, an exponential stellar disc, a stellar bulge (except for one model) and a thin gaseous disc, constructed to be in self gravitating equilibrium if evolved autonomously \citep{springel05b}. The disc galaxies are initially set up with 10$^5$--10$^6$ particles for the dark matter halo, 22,938--51,250 particles for the stellar component, and 7,688--25,625 particles for the gas. The dark matter haloes have concentration parameters related to their total masses and condensation redshifts according to \citet{bullock01}, {implying that for a fixed mass the halo concentration increases with redshift}. The total mass is related to the virial velocity $V_{\rm vir}$ and the Hubble constant $H(z)$ at redshift $z$ as $M_{\rm vir}=V_{\rm vir}^3/[10GH(z)]$. For all galaxies, the baryonic disc is constituted by a gaseous and stellar component, having a mass fraction $m_{\rm d}=0.041$ of the total mass, while the bulge (when included) consists of a stellar component only, having a mass fraction $m_{\rm b}=0.008$ of the total mass. {These mass fractions are chosen to be consistent with the fiducial values from recent literature \citep[e.g.][]{springel05b} and are based on the original constraint of $0.03<m_{\rm d}<0.05$ by \citet{mo98}.} The fraction of total angular momentum that is constituted by the disc ($j_{\rm d}$) is taken identical to $m_d$. The scale-length of the bulge and the vertical scale-length of the disc are 0.2 times the radial scale-length of the disc, which is determined by the {degree of rotation \citep{mo98} through the spin parameter $\lambda\equiv J |E| / GM_{\rm vir}^{5/2}$, in which $J$ is the angular momentum of the halo and $E$ its total energy}. Table~\ref{tab:discs} lists the remaining properties for the various model runs, i.e. the gas fraction of the baryonic disc $f_{\rm gas}$, the total mass $M_{\rm vir}$, the spin parameter $\lambda$, the number of particles in the different components of the model galaxies, and the particle masses of the halo particles $M_{\rm part}^{\rm halo}$ and baryonic particles $M_{\rm part}^{\rm bary}$.

\begin{table*}\centering
\caption[]{\label{tab:mergers}
     Details of the initial conditions for the galaxy merger models.}
\begin{tabular}{c c c r r r r r c}
\hline
ID & Progenitors & Mass ratio & $\theta_1$ & $\phi_1$ & $\theta_2$ & $\phi_2$ & ${R_{\rm peri}}^a$ & Comments
\\\hline
1m1      &    1dA/1dA    &      1:1     &      0         &     0     &      0         &     0       &       6      &     prograde-prograde (PP)      \\
1m2      &    1dB/1dB    &      1:1     &      0         &     0     &      0         &     0       &       6      &     PP        \\
1m3      &    1dC/1dC    &      1:1     &      0        &      0    &       0        &     0        &      6      &     PP      \\
1m4      &    1dB/1dD    &      1:2     &      0         &     0     &      0         &     0       &       6      &     PP          \\
1m5    &    1dE/1dE     &      1:1     &      0        &     0     &      0         &     0        &      6      &     PP           \\
1m6    &    1dF/1dF     &      1:1     &       0        &     0     &      0         &     0        &      6      &     PP           \\
1m7    &    1dG/1dG     &      1:1     &      0        &     0     &      0         &     0        &      6      &     PP           \\
1m8    &    1dH/1dH     &      1:1     &      0        &     0     &      0         &     0        &      6      &     PP           \\
1m9    &    1dI/1dI     &      1:1     &      0        &     0     &      0         &     0        &      6      &     PP           \\\hline
1m10      &    1dB/1dB    &      1:1     &      60       &     45   &     -45       &    -30    &      6      &     PP inclined/near-polar           \\
1m11      &    1dB/1dB    &      1:1     &      180     &     0     &      0         &     0        &      6      &     prograde-retrograde (PR)           \\
1m12      &    1dB/1dB    &      1:1     &      -120   &     45   &      -45      &     -30    &      6      &     PR inclined/near-polar           \\
1m13      &    1dB/1dB    &      1:1     &      180     &     0     &      180     &     0       &      6      &     retrograde-retrograde (RR)          \\
1m14      &    1dB/1dB    &      1:1     &      -120   &     45   &      135     &     -30    &      6      &     RR inclined/near-polar           \\
1m15    &    1dB/1dB     &      1:1     &      0        &     0     &      0         &     0        &      12    &     wide PP           \\
1m16    &    1dC/1dG     &      1:1     &      -120  &     45   &      -45     &     -30    &      10    &     PR inclined/near-polar           \\\hline
1m17      &    1dB/1dB    &      1:1     &      0       &     0   &     71       &    30    &      6      &     `Barnes'           \\
1m18      &    1dB/1dB    &      1:1     &      -109       &     90   &     71       &    90   &      6      &     `Barnes'           \\
1m19      &    1dB/1dB    &      1:1     &      -109       &     -30   &     71       &    -30    &      6      &     `Barnes'           \\
1m20      &    1dB/1dB    &      1:1     &      -109       &     30   &     180       &    0    &      6      &     `Barnes'           \\
1m21      &    1dB/1dB    &      1:1     &      0       &     0   &     71       &    90    &      6      &     `Barnes'           \\
1m22      &    1dB/1dB    &      1:1     &      -109       &     -30   &     71       &    30    &      6      &     `Barnes'           \\
1m23      &    1dB/1dB    &      1:1     &      -109       &     30   &     71       &    -30    &      6      &     `Barnes'           \\
1m24      &    1dB/1dB    &      1:1     &      -109       &     90   &     180       &    0    &      6      &     `Barnes'           \\\hline
\multicolumn{9}{l}{$^a$In kpc. All angles are in degrees.}
\end{tabular} 
\end{table*}
The gas fractions of the galaxy models are chosen to cover the range from typical star forming galaxies. Most of the total masses represent Milky Way type galaxies, with the two exceptions being one half and one tenth of that mass, enabling simulations of unequal-mass major and minor mergers. {The parameter $\lambda$ represents the degree of rotational support, and} is set in accordance with typical spiral galaxies in cosmological simulations \citep[$\langle\lambda\rangle=0.045$,][]{vitvitska02}, except in one case, where we evaluate the influence of the radial scale-length of the disc on the cluster population. The number of halo particles is chosen to ensure sufficient smoothing of the dark matter halo, and the number of stellar disc, stellar bulge and gas particles are chosen to minimise the mass difference between the particles and alleviate two-body effects.  

The model runs for galaxy mergers are initialised by positioning two disc galaxy models on Keplerian parabolic orbital trajectories\footnote{About 50\% of the mergers in cosmological simulations are on (near-)parabolic orbits \citep{khochfar06b}.} with initial separations of approximately 200~kpc. The actual orbit will decay due to dynamical friction, which leads to the merging of the galaxies. The orbital geometry of an interaction is characterised by the directions of the angular momentum vectors of the two galaxy discs and the pericentre distance of the parabolic orbit $R_{\rm peri}$. The angular momentum vectors of the galaxies are determined in spherical coordinates by angles $\theta$ (rotation perpendicular to the orbital plane) and $\phi$ (rotation in the orbital plane). These and other relevant parameters are listed in Table~\ref{tab:mergers}, where the different model runs are summarised.

The initial conditions in Table~\ref{tab:mergers} are divided in three categories. The first set of eight runs follow a common configuration, in which the discs rotate in the orbital plane. They are used to test the influence of the gas fraction and mass ratio of the progenitor discs, and of additional progenitor disc properties such as the presence of a bulge, total galaxy mass, and spin parameter (or the radial scale-length of the disc). The initial conditions for the six subsequent runs are constructed to assess the impact of orbital parameters on the star cluster population. We rotate the progenitor discs to include retrograde rotation and near-polar orbits, which represent the opposite extreme with respect to the co-planar configurations of the first eight runs. The effect of a wider orbit (a larger pericentre distance) is also considered, and a `random' major merger is also included, in which two progenitors with different spin parameters are placed on a near-polar, prograde-retrograde orbit. The third group contains the eight `random' configurations from \citet{hopkins09} \citep[see][]{barnes88}, which all have equal probabilities of occurring in nature.

All galaxies are generated without any star clusters, and we set $t=0$ after 300~Myr of evolution to initialise the cluster population. As described in Sects.~\ref{sec:clform} and~\ref{sec:clevo}, the clusters have masses between 10$^2$ and $\sim10^{5.8}~\msun$, following a \citet{schechter76} type initial mass function. The chemical composition of the clusters is set to solar metallicity, and we assume a King parameter of $W_0=5$.

The properties of the simulated disc galaxies and galaxy mergers are not intended to cover all of parameter space, but instead should provide a first indication of how the modeled star cluster populations are affected by their galactic environment. This set of simulations represent a basic library that can be used to predict certain characteristics of star cluster populations and to see how well the simulated cluster populations compare to observations.

\section{Isolated disc galaxies} \label{sec:discs}
As a first application of the model, we consider the simulations of the isolated disc galaxies from Table~\ref{tab:discs}. As discussed in Sect.~\ref{sec:clform}, the cluster populations are simulated down to a lower mass limit of 100~\msun, which for our assumed cluster formation efficiency and for the typical cluster formation and disruption rates of disc galaxies yields about 1---$3\times10^5$ clusters per galaxy \citep[also see][]{kruijssen10}. Below, we discuss the mechanisms driving the evolution of individual clusters, and show a number of key properties of the entire cluster population.

\subsection{The evolution of individual star clusters in disc galaxies} \label{sec:trackdisc}
\begin{figure*}
\resizebox{14cm}{!}{\includegraphics[trim=22 0 -22 0]{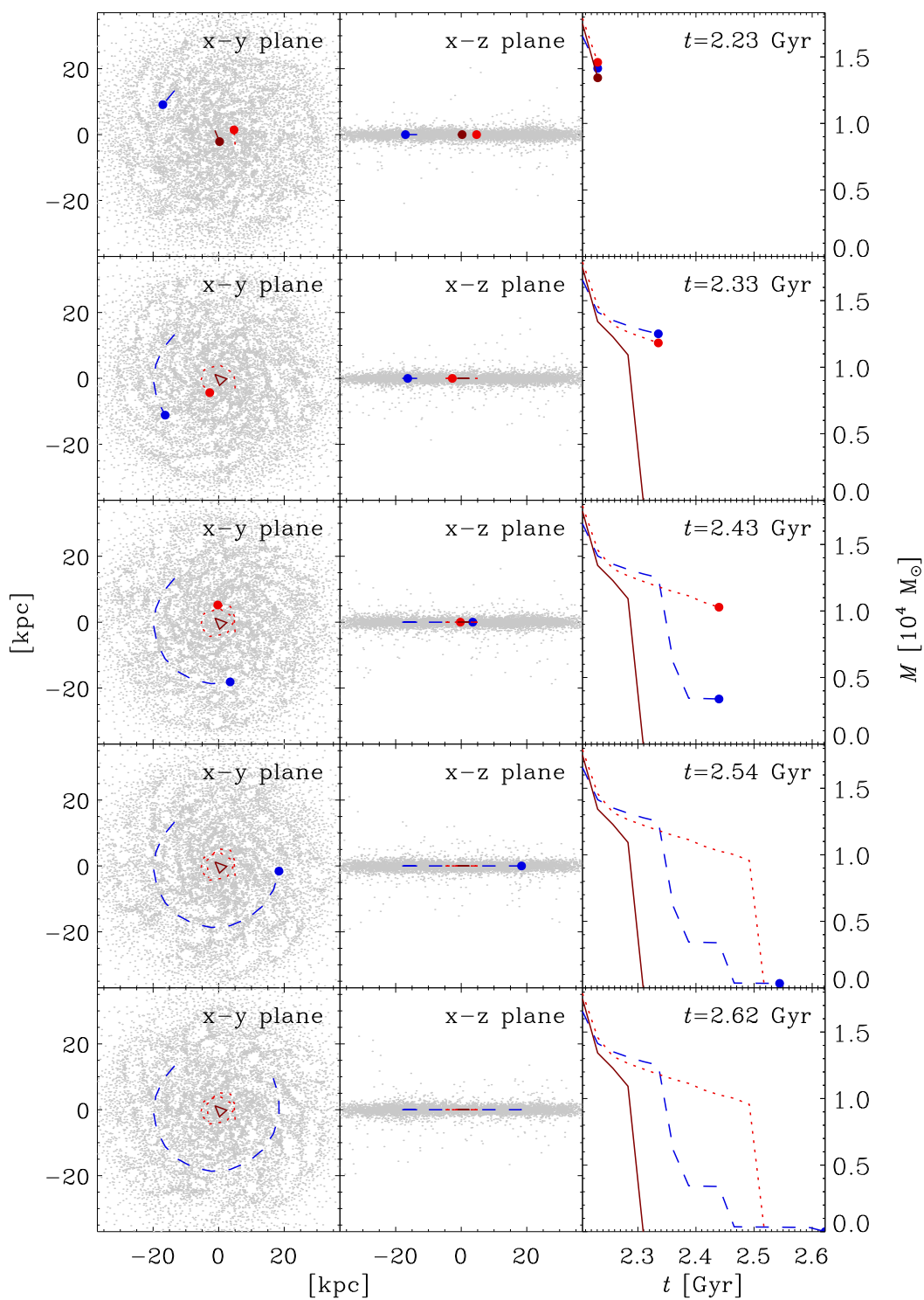}}
\caption[]{\label{fig:trackdisc}
Evolution of the orbits and masses of three clusters orbiting at different galactocentric radii in isolated disc galaxy simulation 1dB. From top to bottom, the consecutive panels show the situation at different times $t$, while from left to right the orbital evolution in the x-y plane (face-on), the orbital evolution in the x-z plane (edge-on), and the mass evolution are shown. The innermost cluster is represented by the dark red solid lines, the middle cluster by red dotted lines, and the outermost cluster by blue dashed lines. If at any particular snapshot a cluster is still undisrupted, its position and mass are marked with thick dots. The orbital trajectories remain visible after the clusters are disrupted. The small grey dots in the x-y and x-z plane views map the distribution of the gas particles in the simulation.
       }
\end{figure*}
To illustrate the effects of disruption due to two-body relaxation and tidal shocks, in Fig.~\ref{fig:trackdisc} we show the orbits and the mass evolution of three star clusters with similar initial masses ($M_{\rm i}\sim 1.8\times 10^4~\msun$) and times of formation ($t\sim 2.20$~Gyr) from simulation 1dB. They are on different orbits and therefore experience differing tidal evolution. Clusters orbiting at small galactocentric radii evolve in a stronger tidal field (and generally have smaller {Jacobi} radii) than clusters orbiting at large galactocentric radii. Also, the number and intensity of tidal shocks is typically larger for clusters orbiting close to the galactic centre, due to the higher gas density in their environment. These differences result in contrasting mass loss histories and total disruption times, as the innermost cluster survives for about 100~Myr, the middle cluster persists for about 300~Myr, and the outer cluster is disrupted after 400~Myr, even though it has the lowest initial mass of the three clusters. The jumps in the mass loss history indicate the effect of tidal shocks, which are generally stronger (and potentially more frequent) for clusters on narrow orbits. Despite these trends, Fig.~\ref{fig:trackdisc} also shows that a cluster can be disrupted by a single tidal shock almost anywhere in the galaxy, even at radii well beyond the solar galactocentric radius. This was also discussed by \citet{gieles06}, who showed that clusters with masses $\la 2\times 10^4~\msun$ can be disrupted during a single encounter with a spatially extended GMC of $10^6~\msun$. Relative to disruption due to subsequent, small encounters, disruption by a single, violent GMC encounter is most prominent for cluster masses of about $10^4~\msun$ \citep{gieles06}. The clusters in Fig.~\ref{fig:trackdisc} are characteristic examples of this. In general, the strongest tidal shocks occur at times when the clusters cross regions of high gas density, for instance during spiral arm passages. This is best seen in the snapshots at $t=2.33$~Gyr and $t=2.43$~Gyr, because between those snapshots the outer cluster is overtaken by a dense region\footnote{The outer cluster is situated beyond the co-rotation radius of the galaxy.}, causing it to lose almost 75\% of its mass due to the rapid change of the tidal field.

\begin{figure}
\resizebox{\hsize}{!}{\includegraphics{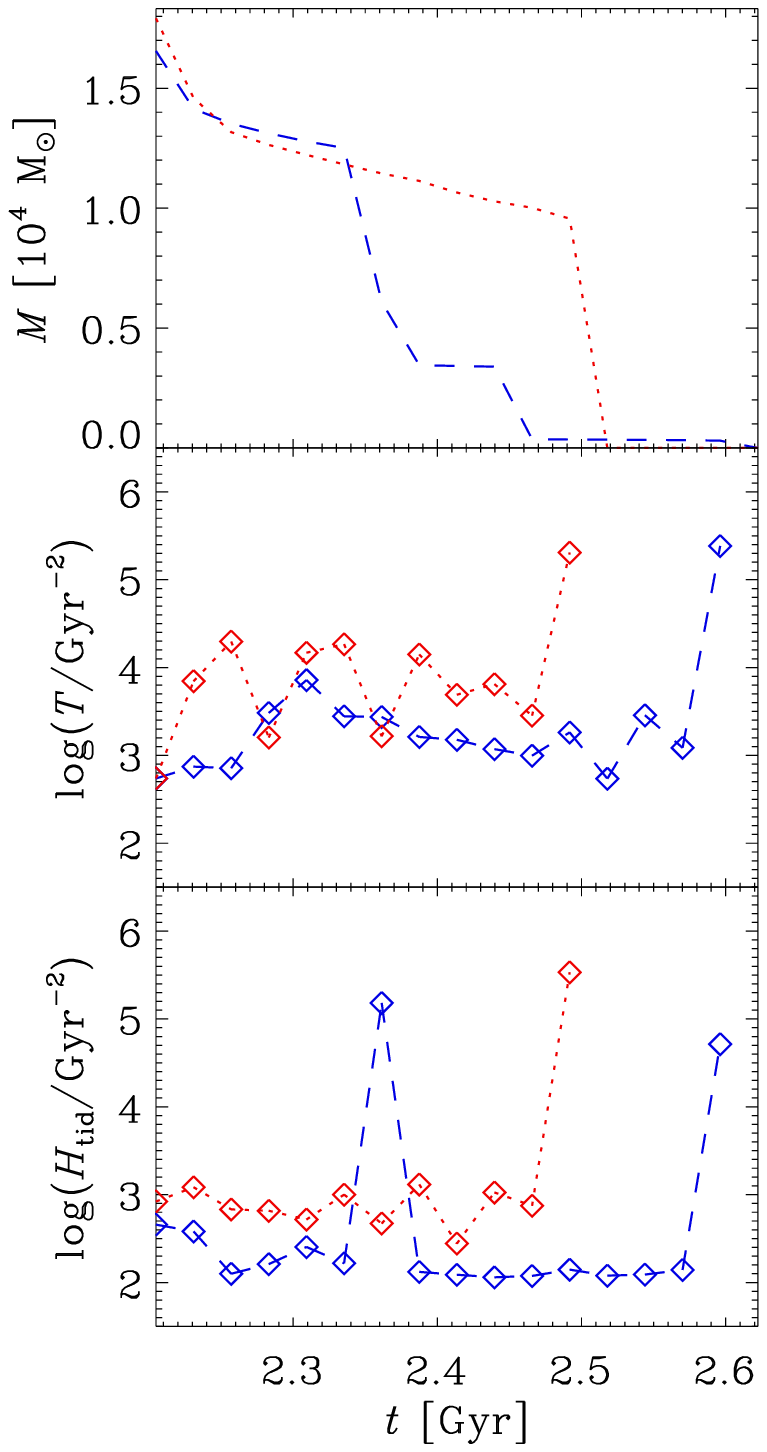}}
\caption[]{\label{fig:tidedisc}
Evolution of the cluster mass and the tidal field for the two outer clusters from Fig.~\ref{fig:trackdisc}, indicated by the same colours and line styles. The diamonds in the middle and bottom panel indicate the times of each snapshot. {\it Top}: The mass evolution. {\it Middle}: Evolution of the tidal field strength experienced by each cluster, defined as the largest eigenvalue of Eq.~\ref{eq:tide} (see Sect.~\ref{sec:clevo}). {\it Bottom}: Running integral of the total amount of shock heating experienced by the cluster (see text and Eq.~\ref{eq:htid}).
       }
\end{figure}

Figure~\ref{fig:tidedisc} illustrates the relation between the mass loss and the tidal field for the two long-lived clusters from Fig.~\ref{fig:trackdisc}. It shows the evolution of the cluster mass, together with the tidal field strength as defined in Sect.~\ref{sec:clevo}, and the running integral that is used to compute the tidal heating parameter $I_{\rm tid}$ in Eq.~\ref{eq:itid}, which is defined as:
\begin{equation}
\label{eq:htid}
 H_{\rm tid}(t) = \sum_{i,j}\left(\int_{t_{\rm last}}^{t} T_{ij}\d t\right)^2 ,
\end{equation}
where $t_{\rm last}$ is the time of the last shock and $t$ is the current time. This quantity represents the accumulated tidal heating since the last shock, and resets after each shock is completed. Contrary to $I_{\rm tid}$, it does not include the adiabatic correction. Therefore, it is only a measure for the tidal shock heating imposed by the tidal field and does not contain any information on the response of the cluster experiencing the shock.

The evolution of the tidal field strength for both clusters in Fig.~\ref{fig:tidedisc} shows that it is indeed larger for the cluster orbiting at the smaller galactocentric radius. A comparison of the tidal field strengths experienced by the clusters just after their formation explains why the outer cluster loses its mass more slowly initially. The moments at which both clusters suffer their first violent mass decrease can be associated with jumps in the amount of shock heating, indicating the effect of tidal shocks. In the case of the inner cluster, this gives rise to its total disruption. The second moment of violent mass loss of the outer cluster cannot be coupled with an increase of the shock heating, because the shock took place in between two snapshots and is therefore skipped by the output of the simulation.

The evolution of the clusters in Figs.~\ref{fig:trackdisc} and~\ref{fig:tidedisc} illustrates that the disruption rate varies with time for individual clusters and varies with space when considering the cluster population as a whole. This is very important when interpreting observed star cluster populations. The time-variation of the disruption rate for individual clusters can mask the effect of a different mass dependence of star cluster disruption. For instance, if the disruption parameter $t_0$ in Eq.~\ref{eq:dmdtrlx} were to increase with time because a cluster is leaving a region with a high gas density \citep[also see][]{elmegreen10b}, for that time interval one would derive a lower value of $\gamma$, which is the mass-dependence of the disruption timescale.

When considering the space-variation of the disruption rate throughout a population of clusters, it is inevitable that the mean disruption rate decreases with age\footnote{Provided that the galaxy and formation sites of the clusters do not change much on timescales of $\sim 1$~Gyr. Galaxy mergers are a clear exception to this.}, because clusters in disruptive environments have the shortest lifetimes and never reach old ages, while clusters in less violent settings become older. This process can be regarded as a form of `natural selection' acting on the star cluster population, and tends to flatten the age distributions of star clusters (see Sect.~\ref{sec:agedis}).

The time-variation of the disruption rate is particularly interesting in view of recent discussions in literature, in which it is debated whether or not star cluster disruption depends on the cluster mass \citep[e.g.][]{fall05,whitmore07,gieles08b,larsen09,bastian09}. It is crucial that environmental dependences are taken into account before inferring any conclusions about the mechanisms driving cluster disruption from observations, because the age and mass distributions of clusters are susceptible to variations in the environment. This holds particular relevance in non-equillibrium settings such as interacting galaxies (see Sect.~\ref{sec:mergers}).

\subsection{The variation of the disruption rate with galactocentric radius} \label{sec:agedisc}
A second consequence of the variability of the disruption rate is related to its variation with space. It implies that the properties of the star cluster population, such as the slope of the age and mass distribution will depend on the local environment within a galaxy. Galaxy-wide distributions may indicate the average properties of the cluster population, but interpreting them can yield systematic errors when assuming the disruption rate is the same everywhere in the galaxy. For instance, the effects of cluster disruption are stronger towards the galactic centre than in the outskirts of a galaxy, implying that the properties of the cluster populations in both regions will differ.

\begin{figure}
\resizebox{\hsize}{!}{\includegraphics{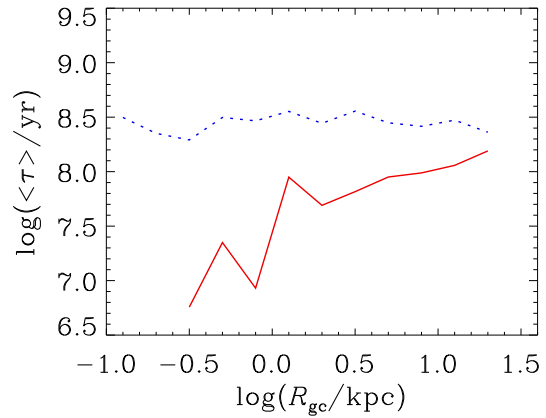}}
\caption[]{\label{fig:agedisc}
Mean age of star clusters $\langle\tau\rangle$ as a function of galactocentric radius $R_{\rm gc}$ for two galaxy simulations at $t=2.5$~Gyr. The red solid line shows the relation for the physically computed disruption rate from Sect.~\ref{sec:clevo}, while the blue dotted line represents the result for a simulation with a fixed disruption rate.
       }
\end{figure}
The environmental dependences on the star cluster population can be qualitatively illustrated by considering two variations of simulation 1dB. Figure~\ref{fig:agedisc} shows the mean cluster age as a function of galactocentric radius for two galaxy disc simulations: one in which the disruption rate is calculated as described in Sect.~\ref{sec:clevo} (model 1dB), and one with a disruption rate that is constant in time and space (using $t_0=2$~Myr, see Eq.~\ref{eq:dmdtrlx}) and does not include tidal shocks (model 1dB$_{\rm fix}$). For the galaxy with the physically motivated disruption rate, the spatial distribution of the mean cluster age is as expected. The youngest clusters are found in the galactic centre, where the star formation rate density is highest. Due to the high gas density, clusters in the galactic centre disrupt on shorter timescales than in the outskirts of the galaxy, resulting in a mean cluster age that increases with galactocentric radius. This contrasts with the age profile of the cluster population in the galaxy with a fixed disruption rate, where the mean cluster age is approximately constant throughout the galaxy and the scatter is strictly due to local variations in the star formation history and stochastical effects\footnote{{The spatial variation of the star formation rate (SFR) cannot produce the behaviour of the mean cluster age that is shown in Fig.~\ref{fig:agedisc}. Without a time-variation, the cluster age distributions at different galactocentric radii would still yield the same mean age, irrespective of the relative formation rates. Within a stable disc galaxy, the relative time variations at different galactocentric radii are not sufficiently large nor persistent enough to cause a spatial trend of the mean cluster age, as is also shown by the line denoting the galaxy with a fixed disruption rate. Indeed, Fig.~\ref{fig:agedisc} could have been made at any time in our simulation other than the time shown, and the mean age would have shown the same spatial variation.}}. Observations of cluster populations in real galaxies \citep[e.g.][]{vandenbergh80,gieles05a,froebrich10} show that the mean age increases with galactocentric radius, contrary to the result for a fixed disruption rate. For the inner disc of M51, \citet{gieles05a} find that the disruption rate varies by a factor 1.8 between radial intervals of 1--3~kpc and 3--5~kpc. Assuming that the mean age is directly proportional to the disruption timescale, this is of the same order of magnitude as for the model shown in Fig.~\ref{fig:agedisc}, for which we find that the ratio between the mean ages of the clusters in these intervals is 1.4. 

These results substantiate that star cluster disruption is indeed driven by environmental effects. Additionally, they show that the suggestion of a disruption rate that increases with the star formation rate, which is found when considering variations between different galaxies \citep[e.g.][]{boutloukos03,lamers05a}, also holds within a single galaxy. This is easily understood by noting that both the formation and disruption of clusters peak in dense environments.

\subsection{The relative importance of tidal shocks and two-body relaxation}
\begin{figure}
\resizebox{\hsize}{!}{\includegraphics{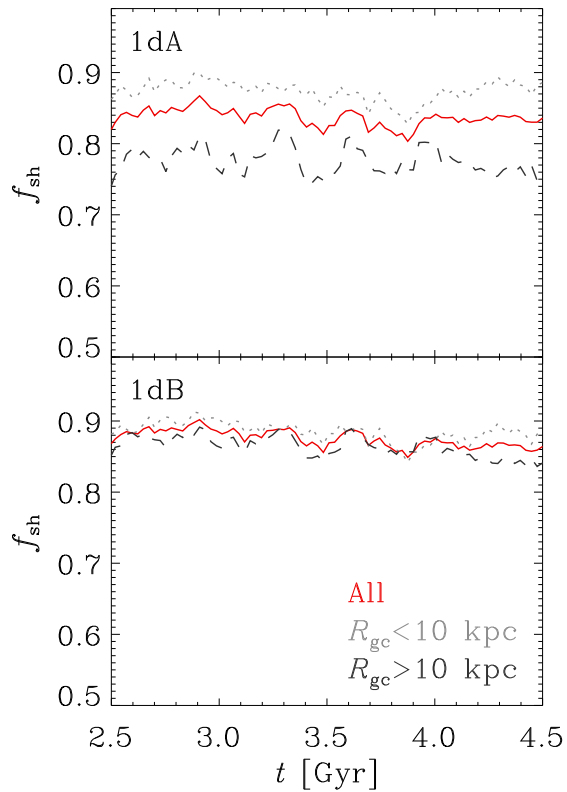}}
\caption[]{\label{fig:rlxsh}
Fraction of disruption due to tidal shocks as a function of time. The red solid line shows the evolution for all clusters, while the light grey dotted line only includes clusters within a galactocentric radius of 10~kpc, and the dark grey dashed line represents the evolution for the clusters beyond 10~kpc. {\it Top}: disc galaxy with a gas fraction $f_{\rm gas}=0.20$ (simulation 1dA). {\it Bottom}: disc galaxy with a gas fraction $f_{\rm gas}=0.30$ (simulation 1dB).
       }
\end{figure}
The relative contributions to star cluster disruption of two-body relaxation and tidal shocks can be quantified by considering the number of clusters in simulations for which either mechanism is neglected. The fraction of the total disruption contributed by tidal shocks $f_{\rm sh}$ is then given by
\begin{equation}
\label{eq:fsh}
f_{\rm sh}\equiv1-\frac{N_{\rm both}}{N_{\rm rlx}} ,
\end{equation}
where $N_{\rm both}$ is the number of clusters in a simulation including both disruption mechanisms (e.g. 1dA and 1dB), and $N_{\rm rlx}$ is the number of clusters in a simulation for which only disruption by two-body relaxation is included and tidal shocks are not considered (e.g. 1dA$_{\rm rlx}$ and 1dB$_{\rm rlx}$). Per definition $N_{\rm both}<N_{\rm rlx}$. For different radial bins, $f_{\rm sh}$ is shown as a function of time in the top panel of Fig.~\ref{fig:rlxsh} for two galaxies with different gas fractions (1dA and 1dB). The contribution by tidal shocks is typically 80--85\% of all disruption, which is very similar to the analytical estimate by \citet{lamers06a} for the solar neighbourhood. The value increases with the gas fraction of the disc, because GMCs and spiral arms are the most important sources of tidal shocks. For relatively gas-rich discs such as in simulation 1dB ($f_{\rm gas}=0.30$), the contribution from shocks does not vary much with galactocentric radius, but for gas-poorer discs, shocks are more important in the inner regions of the disc. This occurs because beyond a certain galactocentric radius the gas density becomes too low to sustain star formation \citep[e.g][]{kennicutt89,schaye04,pelupessy04}, yielding less or no energy injection by feedback and a less filamentary gas distribution, which in turn implies that tidal shocks are less important. The characteristic radius for this transition is smaller in gas-poor galaxies, which is illustrated by the contrast between the inner and outer parts of the disc in the upper panel Fig.~\ref{fig:rlxsh}. Because discs also become more gas-poor as they age, the relative contribution by shocks slightly decreases with time.

For the adopted mass-radius relation, the ratio between the disruption timescales due to two-body relaxation and tidal shocks is $t_{\rm rlx}/t_{\rm sh}\propto M^{0.3}I_{\rm tid}$. If this ratio is larger than unity, tidal shocks dominate cluster disruption, while a ratio below unity implies that disruption is mainly driven by two-body relaxation. The relative importance of tidal shocks increases with cluster mass until a few times 10$^4$~\msun, when the adiabatic correction in the tidal heating parameter $I_{\rm tid}$ (see Eq.~\ref{eq:itid}) becomes non-negligible and inhibits disruption due to tidal shocks. This means that the relative importance of tidal shocks peaks at a certain cluster mass. For the parameters in this paper, this is $f_{\rm sh}\approx 0.9$ at $M\sim 10^4~\msun$, but the precise value depends on the mass-radius relation.

\subsection{The age distributions of star clusters in disc galaxies} \label{sec:agedis}
The balance between cluster formation and destruction gives rise to a cluster population with a certain age distribution. The age distribution of star clusters is often used as a probe to study star cluster disruption \citep[e.g.][]{gieles05a,chandar06}, or to assess the formation history of a galaxy \citep{hunter03,gieles05a,smith07}. In order to obtain a reliable interpretation of the cluster age distribution, it is important to understand its evolution in different galaxies.

To investigate possible correlations between the cluster age distribution and galaxy properties, we have fitted the logarithmic slope $\alpha$ of the age distribution (d$N$/d$\tau\propto \tau^\alpha$) in the age range $\log{(\tau/{\rm yr})}=7.7$--9 for all snapshots of our galaxy disc simulations. When constructing the age distribution, we consider all available clusters, implying that the samples are mass-limited with $M\geq 100~\msun$. We have also fitted the logarithmic slope $\beta$ of the SFR-corrected age distributions in that range ([d$N$/d$\tau]/{\rm SFR}\propto \tau^\beta$). The age range has been chosen such that the effects of gas expulsion due to supernovae are no longer relevant and a sufficiently large part of the age distribution is covered to obtain a reliable slope. The fits have been made with 13 bins in the specified age range, using a variable bin width to accommodate equal numbers of clusters in each bin. The clusters outside the fitted age range are binned using the same number of clusters per bin. We have adopted Poissonian errors for d$N$/d$\tau$, scaling the square root of the number of parent star particles instead of using the number of clusters in each bin, because the ages of the clusters within a single star particle are identical (see Sect.~\ref{sec:clform}). In practice, this means that the relative error decreases with age, because the mean number of clusters per particle decreases. To ensure a reliable fit, the slopes have only been measured at times when a galaxy contains clusters older than 1~Gyr.

\begin{figure}
\resizebox{\hsize}{!}{\includegraphics{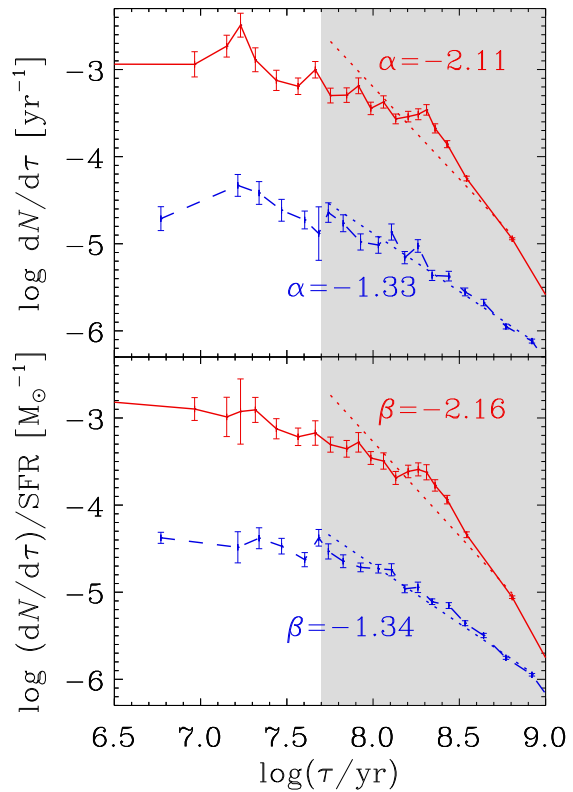}}
\caption[]{\label{fig:agedis}
{\it Top}: Age distributions of clusters in simulations 1dB (red solid line) and 1dG (blue dashed line, vertically offset by 1.5~dex) at $t=3$~Gyr for ages in the range $\log{(\tau/{\rm yr})}=7.7$--9. The dotted lines represent power law fits to the data in the age range indicated by the shaded area. {\it Bottom}: Same as above, but with the age distributions divided by the star formation rate (SFR-corrected). The error bars are computed as described in the text. In both panels the fitted slopes are indicated.
       }
\end{figure}

Given the time interval between subsequent output snapshots, the above procedure results in 1175 fitted age distribution slopes, covering eight different disc galaxy models. These slopes should be considered `mean' slopes for the specified age range, because the age distribution does not always follow a single logarithmic slope over the fitted age range. This is illustrated in Fig.~\ref{fig:agedis}, which shows (SFR-(un)corrected) age distributions for two different galaxies, of which the upper one (1dB) is indeed ill-fitted by a single power law. For the SFR-corrected age distributions, the negative slope is solely the result of disruption, with small variations due to stochastical effects. The SFR does not vary much in isolated galaxies, and therefore only affects the fitted slopes by a few hundredths. 

In models with the same disruption rate for all clusters and a constant SFR, the `classical' age distribution is characterised by two components \citep[e.g.][]{boutloukos03,lamers05}. At young ages, the age distribution is flat, because no clusters are disrupted within such a short time interval. Beyond the lifetime of the lowest mass cluster, the age distribution steepens. This is the disruptive (old) end of the age distribution, which has a slope of $\beta=-1/\gamma$, where $\gamma$ is the mass dependence of the disruption timescale (see Eq.~\ref{eq:dmdtrlx})\footnote{It is assumed that the logarithmic slope of the cluster initial mass function (d$N$/d$M$) is $-2$.}.

The models presented in this paper assume a more realistic formulation, in which the disruption rate is affected by the variation of the tidal field strength and by tidal shocks. Nonetheless, for the sake of illustration it is important to indicate what the disruption-dominated slope of the age distribution would be if the tidal field strength would be the same throughout a galaxy, and all clusters would experience the same tidal shocks. In such a scenario, the adopted value of $\gamma=0.62$ for disruption due to two-body relaxation gives $\beta=-1.61$. For rapid shocks (i.e. a negligible adiabatic expansion during the shock), our adopted mass-radius relation yields $\beta=-3.08$, while for slow shocks (i.e. a dominant adiabatic expansion during the shock) we have $\beta=-1.23$. Fast disruption (i.e. rapid shocks) thus yields a steeper age distribution than slow disruption.

Evidently, the bulk properties of the cluster population are determined by a combination of the above mechanisms, covering a range of tidal field and shock strengths. As such, the fitted slope of the age distribution is not only determined by the mass dependence of the disruption timescale, but also by possible trends of the disruption rate with cluster age and by the rapidity of disruption in general.

The difference between the age distributions shown in Fig.~\ref{fig:agedis} should be the result of the differences between the initial conditions of both simulations. Galaxy 1dG only differs from 1dB by its (larger) spin parameter (see Table~\ref{tab:discs}), implying a correspondingly larger scale radius and lower (gas) density, which yields a lower disruption rate (see Sects.~\ref{sec:trackdisc} and~\ref{sec:agedisc}). Because of the more rapid disruption in simulation 1dB only very few clusters survive for $\sim 1$~Gyr, causing the depletion in the oldest bin, which in turn steepens the fitted slope. Again, faster disruption implies less survivors and a potentially steeper fitted slope than for slow disruption.

\begin{figure}
\resizebox{\hsize}{!}{\includegraphics{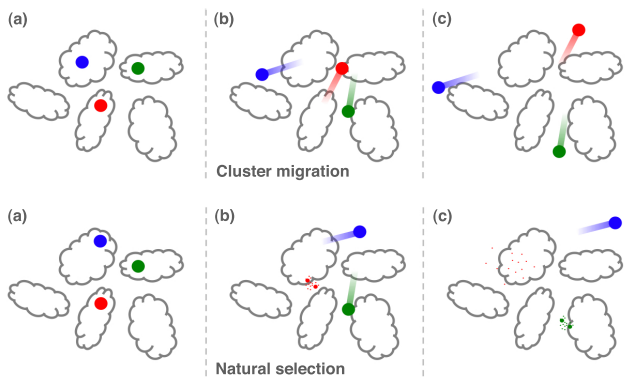}}
\caption[]{\label{fig:migrationselection}
Schematic representation of the two processes leading to a cluster disruption rate that decreases with age. {\it Top}: Cluster migration. {\it Bottom}: Natural selection. The large dots mark star clusters, the small dots represent debris from disrupted clusters, and the clouds denote gas clouds. Time increases from left to right in both sequences.
    }
\end{figure}
Another effect is that the disruption rate due to tidal shocks will typically decrease as clusters age. This happens for two reasons (also see Fig.~\ref{fig:migrationselection}):
\begin{itemize}
\item[(1)] `Cluster migration': because clusters move out of their primordial environment, the ambient gas density typically decreases as they age, giving rise to fewer tidal shocks and a lower disruption rate at older ages \citep[see][]{elmegreen10b}. This evolution of the mean disruption rate is more pronounced if the density contrast between the star forming region and its surroundings is large.
\item[(2)] `Natural selection': at any given time, clusters in regions with a high disruption rate are less likely to survive than clusters in low disruption rate regions. Such selection implies that at older ages only the clusters in low disruption rate regions are left, causing the disruption rate to decrease with age (also see Sect.~\ref{sec:trackdisc} and Fig.~\ref{fig:trackdisc}). This evolution of the mean disruption rate is more pronounced if there is a large spread in disruption rates, like in galaxies with large density contrasts between different regions.
\end{itemize}
These two effects make the disruptive end of the age distribution shallower and steepen the young end of the age distribution. In the extreme case, this can lead to an age distribution following a single power law with a slope of $-1$ over the majority of the age range. The effects of cluster migration and natural selection are strongest for galaxies with low gas densities, because in those galaxies the density contrast between star forming regions and their surroundings is larger than in high gas density galaxies per definition\footnote{This holds for isolated galaxies.}. While already present in simulation 1dG, it could thus be even more important in dwarf galaxies, which have very low gas densities. For out-of-equillibrium systems such as galaxy mergers, the dependence of the age distribution on the mean gas density is different (see Sect.~\ref{sec:mergers}).

Above, we discussed: (1) the different disruption processes shaping the age distribution, (2) the effect of the largest possible cluster lifetime on the fitted slope, (3) the effect of cluster migration, and (4) the effect of natural selection. For all four of those, galactic environments with high gas densities steepen the slope. As discussed at length before, cluster disruption is governed by the gas density ($\rho_{\rm gas}$), implying that in isolated disc galaxies, the fitted slope of the age distribution can be used as a measure for the rapidity of cluster disruption. The gas density also sets the star formation rate density ($\rho_{\rm SFR}$) of a galaxy through the Schmidt-Kennicutt law \citep{schmidt59,kennicutt98b}. One would therefore expect a correlation between the fitted slopes of the age distributions in different galaxies and their star formation rate density. To obtain a measure for the star formation rate density that can be determined and compared for disc galaxies as well as for galaxy mergers at any time during their interaction, we define the mean star formation rate density within a sphere with a radius equal to the half-mass radius of the gas $R_{\rm h,gas}$:
\begin{equation}
\label{eq:sfrd}
\rho_{\rm h,SFR}\equiv\frac{{\rm SFR}_{\rm h}}{V_{\rm h,gas}}=\frac{3}{4\pi}\frac{{\rm SFR}_{\rm h}}{R_{\rm h,gas}^3} ,
\end{equation}
with $V_{\rm h,gas}$ the volume of the sphere, and SFR$_{\rm h}$ the star formation rate within $V_{\rm h,gas}$. For isolated disc galaxies, most if not all of the star formation occurs within $R_{\rm h,gas}$.
\begin{figure}
\resizebox{\hsize}{!}{\includegraphics{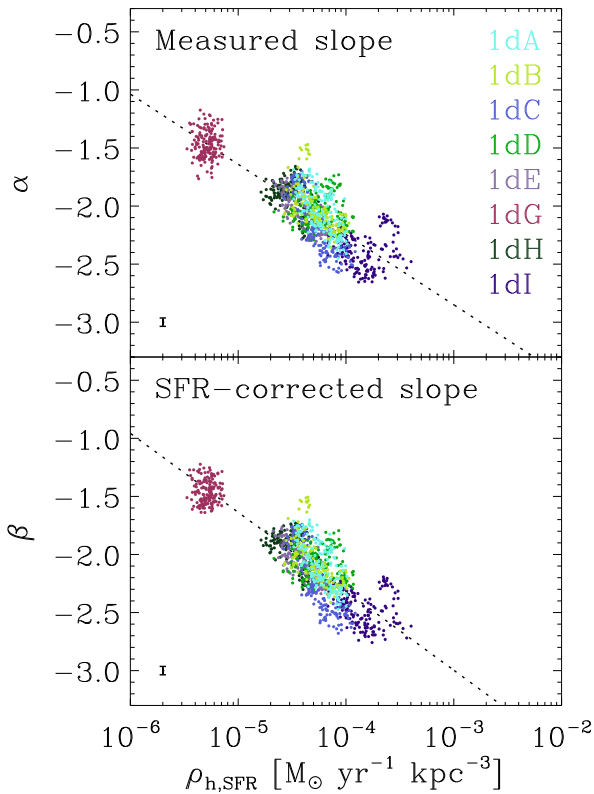}}
\caption[]{\label{fig:dslope}
Relation between the fitted logarithmic slope of the cluster age distribution in the age range $\log{(\tau/{\rm yr})}=7.7$--9 and the mean star formation rate density $\rho_{\rm h,SFR}$, which is defined for a sphere with a radius equal to the half-mass radius of the gas. Each point represents one galaxy snapshot. The snapshots from the different galaxy simulations are colour-coded as indicated in the legend. The best fit to the data is shown as a dotted line, while the typical error on each data point is indicated in the bottom left corner. {\it Top}: showing the measured (unaltered) slopes of the cluster age distributions. {\it Bottom}: showing slopes that are corrected for the variation of the star formation rate (SFR).
       }
\end{figure}

We show the relations between $\rho_{\rm h,SFR}$ and the fitted slope of the cluster age distribution $\alpha$ and fitted slope of the SFR-corrected age distribution $\beta$ in Fig.~\ref{fig:dslope} for all 1175 fits.  As expected, it shows an inverse correlation between the slope of the cluster age distribution and the star formation rate density. For the uncorrected slope $\alpha$, the fitted relation is given by
\begin{equation}
\label{eq:alpha}
\alpha=C-0.60\log{\rho_{\rm h,SFR}} ,
\end{equation}
where $C=-4.66$ is a fitting constant that has no particular physical meaning because we determine $\rho_{\rm h,SFR}$ for a sphere of which a non-negligible fraction is constituted by empty space. If the slopes of the age distributions are corrected for the variation of the SFR instead of using the raw age distributions, we obtain the relation
\begin{equation}
\label{eq:beta}
\beta=C-0.68\log{\rho_{\rm h,SFR}} ,
\end{equation}
with $C=-5.04$. The errors on the fitted slopes in Eqs.~\ref{eq:alpha} and~\ref{eq:beta} are smaller than the listed accuracy. The fitted slopes vary by less than 0.03 if the galaxy in the top-left corner of both panels in Fig.~\ref{fig:dslope} (1dG) is excluded, which underlines the reliability of the fits.

The physical correlation between the slope of the age distribution and the star formation rate density is best described by Eq.~\ref{eq:beta}, because in isolated discs $\beta$ is independent of the variation of the SFR. Conversely, the relation between $\alpha$ and $\rho_{\rm h,SFR}$ (Eq.~\ref{eq:alpha}) would be relevant for comparison with observations. Either way, the implication of both relations is that the rate of cluster disruption increases with the star formation rate density. In \citet{kruijssen10}, we present a similar result for galaxy mergers, in which the number of clusters decreases during a merger despite the large starbursts and corresponding cluster production. The net destruction of clusters is attributed to enhanced cluster disruption that is driven by the high gas density. The analysis of \citet{kruijssen10} does not rely on the cluster age distributions, but instead considers the number of clusters as a function of time. The number of surviving clusters is found to decrease with increasing starburst intensity, which is similar to the relation presented here.

The scatter around the relation between the slope of the age distribution and the star formation rate density is substantial. Within a single galaxy, $\alpha$ and $\beta$ vary by 0.5 at a given star formation rate density, depending on the moment at which the galaxy is observed. Because it is relatively isolated in the displayed plane, galaxy 1dG in Fig.~\ref{fig:dslope} provides a clear illustration of the spread. Recent debates in literature about the mass-dependence of cluster disruption involve differences of a similar magnitude, quoting slopes of $-1$ \citep{whitmore07,chandar10} to $-1.5$ \citep{boutloukos03,silvavilla10b}. As is shown by Fig.~\ref{fig:dslope}, such variations may occur even within a single galaxy. Figure~\ref{fig:dslope} also illustrates that a slope of $-1$ is more likely to occur in galaxies with low star formation rate densities. As such, both sides of the debate show cases that can arise in the framework for star cluster disruption that is presented in this paper\footnote{The starbursts in galaxy mergers are characterised by high star formation rate densities, yet the slope of the cluster age distribution is reported to be $-1$ \citep{whitmore07}, seemingly contradicting Fig.~\ref{fig:dslope} and Eq.~\ref{eq:alpha}. We discuss the inclusion of galaxy mergers in Sect.~\ref{sec:mergers}.}.

\section{Galaxy mergers} \label{sec:mergers}
We now consider the galaxy merger simulations from Table~\ref{tab:mergers}. We discuss the evolution of individual clusters, as well as the evolution of the cluster population as a whole. The section is concluded with a discussion of the cluster population in a merger remnant.

\subsection{The evolution of individual clusters in galaxy mergers} \label{sec:trackmerger}
\begin{figure*}
\resizebox{14cm}{!}{\includegraphics[trim=22 0 -22 0]{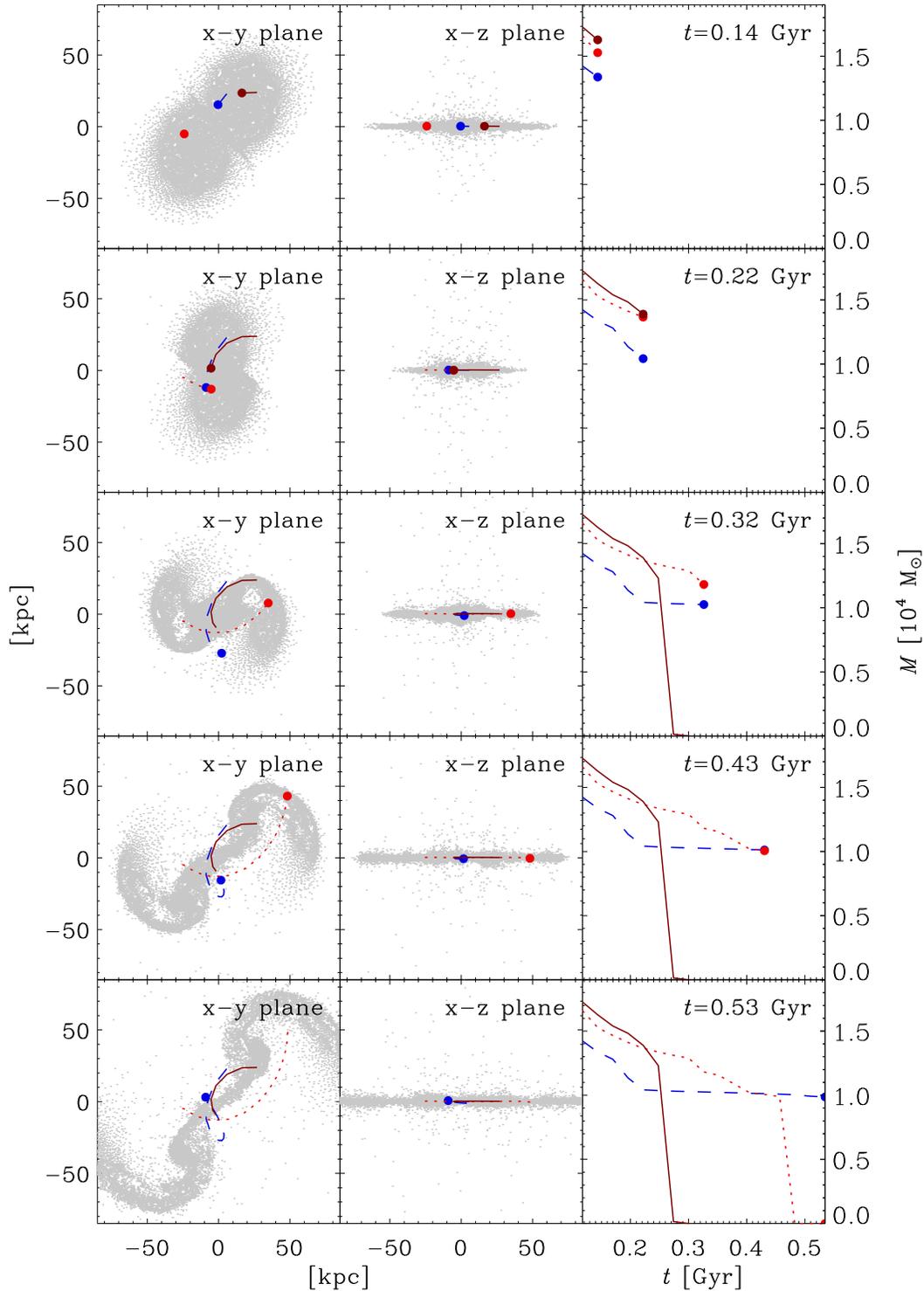}}
\caption[]{\label{fig:trackmerger}
Evolution of the orbits and masses of three clusters in galaxy merger simulation 1m2 during the first pericentre passage of the galaxies. From top to bottom, the consecutive panels show the situation at different times $t$, while from left to right the orbital evolution in the x-y plane (face-on), the orbital evolution in the x-z plane (edge-on), and the mass evolution are shown. The respective clusters are represented by dark red solid lines, red dotted lines, and blue dashed lines. If at any particular snapshot a cluster is still undisrupted, its position and mass are marked with thick dots. The orbital trajectories remain visible after the clusters are disrupted. The small grey dots in the x-y and x-z plane views map the distribution of the gas particles in the simulation.
       }
\end{figure*}
Similar to Fig.~\ref{fig:trackdisc} for disc galaxies, the evolution of the orbits and masses of three `representative' clusters from simulation 1m2 are shown in Fig.~\ref{fig:trackmerger}. As in Sect.~\ref{sec:trackdisc}, the clusters have comparable initial masses ($M_{\rm i}\sim 1.5\times 10^4~\msun$) and times of formation ($t\sim 0.12$~Gyr), and the differences in their evolution are the result of their contrasting orbits in different environments.

The snapshots in Fig.~\ref{fig:trackmerger} follow the merger during the first pericentre passage, when the orbital differences between the clusters are partially conserved. This is not the case during the final coalescence of the two galaxies, when violent relaxation randomises the cluster orbits. Just like in isolated disc galaxies (Fig.~\ref{fig:trackdisc}), the cluster closest to the centre of the galaxy has a low survival chance and is disrupted within $\sim 200$~Myr. The two other clusters survive the first passage of the galaxies and experience different evolutionary scenarios. One is ejected from the disc of its parent galaxy (the red dotted cluster in Fig.~\ref{fig:trackmerger}), together with all the surrounding gas and stars, and ends up in the trailing tidal tail of the galaxy. It has a low velocity with respect to the tidal tail, but it does experience an intermediate tidal shock when entering the tidal arm at $t=0.32$~Gyr, and a strong tidal shock when it hits the densest part at $t=0.45$~Gyr, leading to the disruption of the cluster. The other cluster (blue dashed in Fig.~\ref{fig:trackmerger}) is ejected from the disc as well, but it decouples from the surrounding gas. This occurs because the gas collides with the other galaxy and is shocked, which slows it down to form the bridge between both galaxies. By contrast, the cluster retains a ballistic orbit and becomes part of the stellar halo surrounding the galaxies. As a result, the tidal field strength decreases and the frequency of tidal shocks becomes low, since the cluster is only shocked twice per orbit. The tidal shocks occur when the cluster crosses the bridge or the tidal arm and cause it to lose only a few percent of its mass. Under these conditions, the expected disruption time of the cluster is several gigayears. Even though the cluster mass is only $10^4$~\msun, this could increase to 10~Gyr or more when the tidal arms disperse and the merger consumes the remaining gas, provided that the cluster does not fall back into the central region of the merger. This shows that long-lived constituents of the stellar halo surrounding giant elliptical galaxies are already produced during the first pericentre passage of the progenitor galaxies (see Sect.~\ref{sec:remnant}).

The cluster evolution depicted in Fig.~\ref{fig:trackmerger} illustrates the mechanisms of cluster migration and natural selection that were explained in Sect.~\ref{sec:agedis} and Fig.~\ref{fig:migrationselection}. The cluster that decouples from the gas and is ejected into the stellar halo experiences a disruption rate that decreases as the cluster ages, showing how migration influences the evolution of the cluster population. On the other hand, the cluster that initially resides close to the galactic centre is quickly disrupted by the tidal shock of the first pericentre passage, while the two surviving clusters were situated in less dense environments and therefore survive. This shows how natural selection governs which clusters survive, and that the mean disruption rate of the population decreases with age as clusters in disruptive environments are destroyed.

\begin{figure}
\resizebox{\hsize}{!}{\includegraphics{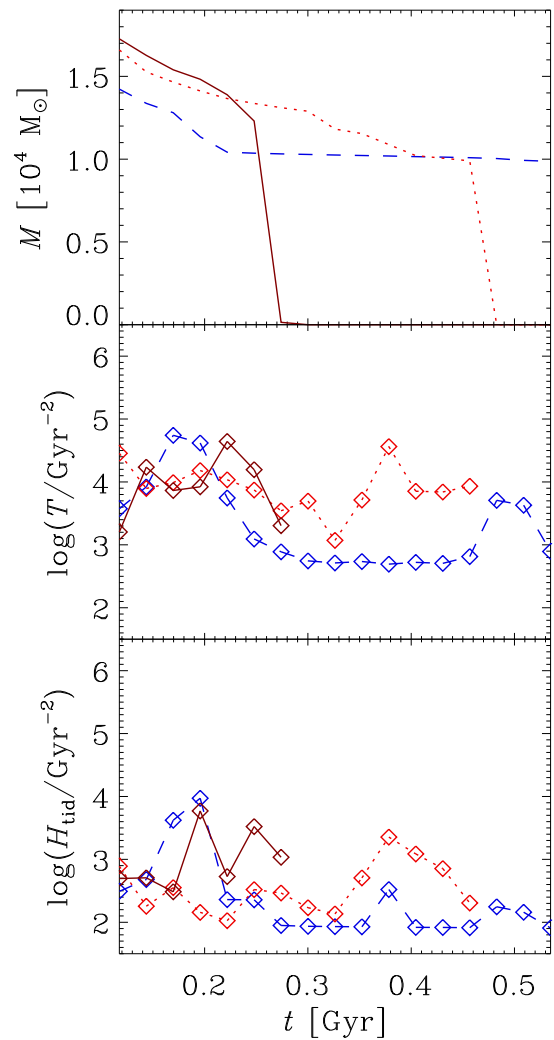}}
\caption[]{\label{fig:tidemerger}
Evolution of the cluster mass and the tidal field for the three clusters from Fig.~\ref{fig:trackmerger}, indicated by the same colours and line styles. The diamonds in the middle and bottom panel indicate the times of each snapshot. {\it Top}: The mass evolution. {\it Middle}: Evolution of the tidal field strength experienced by each cluster, defined as the largest eigenvalue of Eq.~\ref{eq:tide} (see Sect.~\ref{sec:clevo}). {\it Bottom}: Running integral of the total amount of shock heating experienced by the cluster (see Sect.~\ref{sec:trackdisc} and Eq.~\ref{eq:htid}).
       }
\end{figure}
The mass loss histories of the clusters in Fig.~\ref{fig:trackmerger} can be understood by considering the evolution of the tidal field strength and the heating by tidal shocks. Similar to Fig.~\ref{fig:tidedisc} in Sect.~\ref{sec:trackdisc}, this is shown in Fig.~\ref{fig:tidemerger} for the clusters in the merger. It confirms that the short-lived cluster indeed experiences a tidal field strength and tidal shock heating that is only rivaled by the cluster that ends up in the halo. The reason that the halo cluster is not disrupted on the same timescale as the short-lived cluster is that its migration to the halo occurs before disruption would have led to its complete dispersion, thereby decreasing the tidal field strength it experiences. The halo cluster therefore only sustains enhanced disruption when it passes through the bridge between the two galaxies (at $t=0.5$~Gyr), while the short-lived cluster stays in a dense environment and is completely disrupted by two subsequent tidal shocks. By contrast, the cluster in the tidal tail continuously experiences tidal shocks and a stronger tidal field than the halo cluster, because it is moving with the tidal tail and its environment does not change. This leads to an almost constant mass loss rate, which is enhanced by the tidal shocks occurring when the cluster first enters the tidal tail and also when it hits the dense centre of the tail. This second tidal shock occurs in between two snapshots and the corresponding shock heating is therefore not visible in Fig.~\ref{fig:tidemerger}. The decrease of the mean tidal field strength and tidal shock heating with age illustrate the mechanism of natural selection, i.e. the higher survival chances of clusters in quiescent tidal environments. The effects of cluster migration and natural selection are stronger in galaxy mergers than in isolated disc galaxies, because both mechanisms are driven by the variation of the environment with time and space. Such variations are evidently more common in galaxy mergers than in disc galaxies.

\subsection{The age distributions of star clusters in galaxy mergers} \label{sec:agedismerger}
The variation of the environment in galaxy mergers leads to a corresponding evolution of the cluster age distribution. Similar to Sect.~\ref{sec:agedis}, we have fitted the slope of the cluster age distributions in the range $\log{(\tau/{\rm yr})}=7.7$--9 for all galaxy merger simulations, up to the moment of their largest starburst, which typically occurs early on during the final coalescence of both galaxies. The slope is not fitted for later times, because the gas is rapidly consumed during the starburst. At first, this makes the variation of the cluster formation rate dominate the shape of the cluster age distribution, implying that a power law fit is very inaccurate, while later on the age distribution becomes discontinuous due to episodes without any surviving clusters (see Sect.~\ref{sec:remnant} and Fig.~\ref{fig:agedismerger}). Similar to Sect.~\ref{sec:agedis}, we consider all clusters when constructing the age distribution, i.e. the samples are mass-limited with $M\geq 100~\msun$.

\begin{figure}
\resizebox{\hsize}{!}{\includegraphics{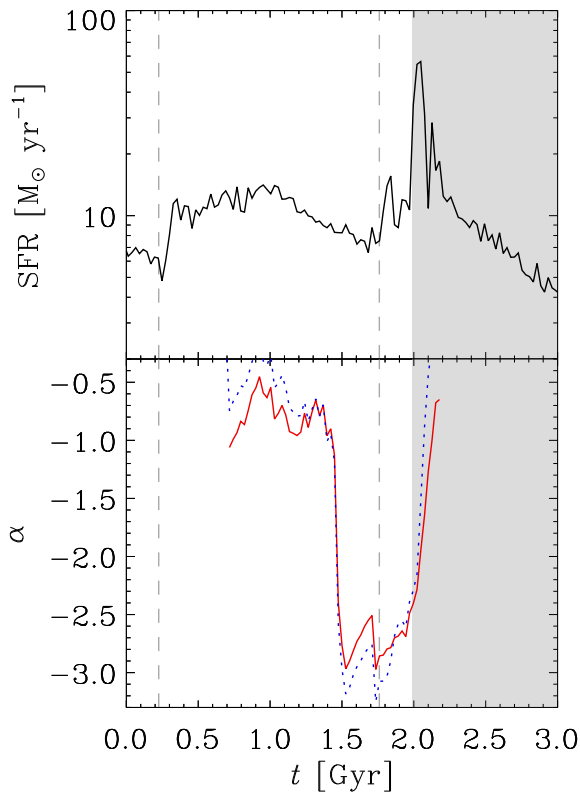}}
\caption[]{\label{fig:slopemerger}
Time evolution of (top) the star formation rate and (bottom) the fitted slope of the age distribution in the range $\log{(\tau/{\rm yr})}=7.7$--9 for merger simulation 1m14, with the red solid line denoting the fit to the actual age distribution $\alpha$, and the blue dotted line denoting the fit to the SFR-corrected age distribution $\beta$. The dashed vertical lines indicate the moments of first and second pericentre passage, and the shaded area marks the period over which the final coalescence occurs.
       }
\end{figure}
In Fig.~\ref{fig:slopemerger}, the star formation history and evolution of the fitted slope of the age distribution are shown for merger simulation 1m14 (see Table~\ref{tab:mergers}). The slope widely changes over the course of the merger, and is shallowest at the times when the star formation rate and star formation rate density are highest, with {typical slopes between $-0.5$ and} $-1$. This behaviour is opposite to what is found in Sect.~\ref{sec:agedis} for isolated disc galaxies, in which the age distribution becomes steeper for higher star formation rate densities. As was discussed in Sect.~\ref{sec:agedis}, a shallow slope indicates that cluster migration and natural selection are important, i.e. that there are large density contrasts in a galaxy, particularly between star forming regions and their surroundings. In isolated disc galaxies, such a large contrast exists for galaxies with an overall low gas density, which then contrasts with the dense star forming regions. This low gas density translates to a low star formation rate density, and gives the relation of Eqs.~\ref{eq:alpha} and~\ref{eq:beta}. In galaxy mergers, the effects of cluster migration and natural selection are largest at the height of the interaction. At that point, the star formation rate (density) peaks, because all gas is funneled to the central regions, leading to a pronounced density contrast between the concentrated star forming volume and the surrounding regions, which hardly contain any gas. In the meanwhile, the ongoing interaction ejects the clusters into the gas-poor stellar halo. The result is visible in Fig.~\ref{fig:slopemerger}, in which the slope of the age distribution evolves to shallower slopes during the starbursts. {The extreme slopes in between the starbursts are typically $-2.5$ to $-3$, which is steeper than found in isolated discs. The reason is illustrated below, in the discussion of Fig.~\ref{fig:agedisburst}.}

Another interesting feature of Fig.~\ref{fig:slopemerger} is the difference between the actual slope $\alpha$ and the SFR-corrected slope $\beta$. Because for $\beta$ the variation of the SFR is divided out, one would expect it to have a more stable evolution than $\alpha$. However, this is not the case in Fig.~\ref{fig:slopemerger}, where the variation of the SFR-corrected slope is larger than that of the actual slope. This is the result of the mechanism identified in \citet{kruijssen10}, who find that the gas density in starbursts is so high that the young clusters formed in the starburst are disrupted on much shorter timescales than in isolated galaxies, even to the extent that the total number of star clusters decreases during a starburst. This counterintuitive result is mainly due to the lowest mass clusters, which are the most numerous for a power law initial mass function with a negative slope\footnote{The index $-2$ of the cluster initial mass function adopted in this study implies that every decade in cluster mass initially has ten times more clusters than the next decade.}. This large number of low mass clusters is susceptible to disruption by the strong tidal shocks in a starburst region. The surprising consequence is that after a certain time interval, the age distribution of all clusters lacks clusters in the age range corresponding to the starburst. After the starburst, the peak in the cluster age distribution shifts to ages just {\it before} the maximum of the starburst (also see Sect.~\ref{sec:remnant} and Fig.~\ref{fig:rgcevo}), when the clusters are still formed in a less violent setting than at the height of the burst, and can be ejected from their primordial regions before the starburst reaches its maximum (like the halo cluster of Fig.~\ref{fig:trackmerger}).

\begin{figure}
\resizebox{\hsize}{!}{\includegraphics{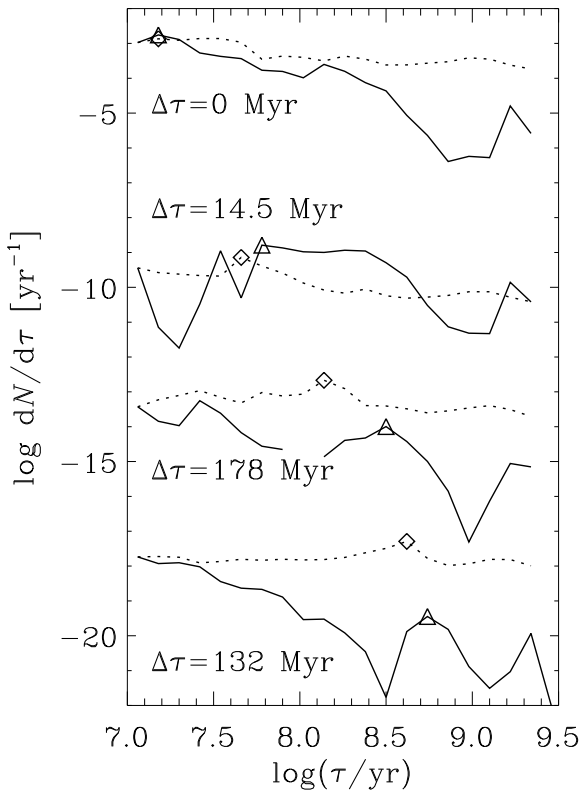}}
\caption[]{\label{fig:agedisburst}
Time-evolution of the cluster age distribution (solid lines) and star formation history (dotted lines) shortly after the second passage of merger simulation 1m14 (at $t\approx 2$~Gyr). From top to bottom, the distributions are shown at times $t=\{2.02,2.07,2.18,2.41\}$~Gyr, corresponding to about \{0,50,150,400\}~Myr after the starburst. For each line, the moment of the starburst is marked with a diamond, while the peak in the cluster age distribution is indicated with a triangle. Each age distribution is shifted down by 5~dex with respect to the distribution above it. The star formation histories are normalised to match the corresponding age distribution at the left end of the lines. For each pair of distributions, the age-offset between the peaks $\Delta\tau$ is indicated.
       }
\end{figure}
The evolution of the age distribution is compared to the star formation history in Fig.~\ref{fig:agedisburst}, which shows the evolution of the age distribution at several times after the major starburst in simulation 1m14. {It illustrates several of the points from the previous paragraphs. The first age distribution (at $t=2.02$~Gyr) shows the cause of the steep slope of about $-3$ just before the second starburst. The fitted slope has steepened relative to isolated galaxies (compare Fig.~\ref{fig:dslope}) due to a deficit of clusters at ages close to $\tau=1$~Gyr, which corresponds to the first starburst, when the high densities triggered enhanced cluster disruption. The same mechanism causes an age-offset between the moment of the second starburst and the peak in the age distribution,} which first emerges when the clusters formed in the starburst have had the time to be disrupted by their environment. This disruption is evident from the minimum in the age distribution at ages slightly younger than the starburst. The exact moment when the offset between the peaks becomes visible depends on the duration and strength of the starburst, but it typically appears 100~Myr after the starburst. The offset grows from $\Delta\tau=14.5$~Myr at $t=2.07$~Gyr to $\Delta\tau=132$~Myr at $t=2.41$~Gyr. As shown in Fig.~\ref{fig:agedisburst}, it is best seen about 150~Myr after the burst. When considering only the massive clusters ($M\ga 10^4$~\msun), which are much less numerous than the low-mass clusters, the deficit of clusters is less prominent. In the extreme case, the offset of the peak in the cluster age distribution with respect to the moment of maximum star formation corresponds to the time interval between the onset and the peak of the starburst.

The age-offset between the starburst and the peak in the cluster age distribution has an interesting consequence in relation to Fig.~\ref{fig:slopemerger}. When dividing the cluster age distribution by the star formation history for a galaxy merger with a recent starburst, the age range corresponding to the starburst will contain even fewer clusters than without the correction for the SFR. As a result, the variation of the age distribution is enhanced with respect to the actual age distribution. This causes the larger variation of $\beta$ than that of $\alpha$ in Fig.~\ref{fig:slopemerger}. The offset between the peaks in the age distributions of the clusters and stars is also seen when considering the formation history of the clusters that survive the merger (see Sect.~\ref{sec:remnant} and Fig.~\ref{fig:rgcevo}), which shows that these clusters are typically formed before instead of during the starburst maximum. It depends on the accuracy of the age determinations of real clusters whether the offset can be distinguished observationally, especially because it is less pronounced for the high cluster masses to which observations are naturally limited.

\begin{figure}
\resizebox{\hsize}{!}{\includegraphics{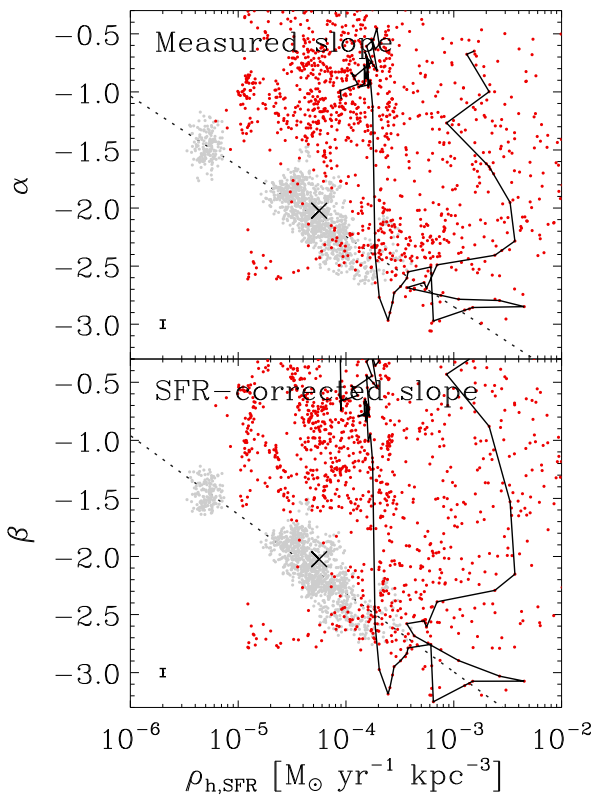}}
\caption[]{\label{fig:agesfhall}
Relation between the fitted logarithmic slope of the cluster age distribution in the age range $\log{(\tau/{\rm yr})}=7.7$--9 and the mean star formation rate density $\rho_{\rm h,SFR}$. Each point represents one galaxy snapshot. The isolated disc galaxies are shown as light grey points, and the galaxy mergers are shown as red points. As in Fig.~\ref{fig:dslope}, the fit to the isolated disc galaxies is represented by a dotted line, while the typical error on each data point is indicated in the bottom left corner. {\it Top}: the measured (unaltered) slopes of the cluster age distributions. {\it Bottom}: slopes that are corrected for the variation of the star formation rate (SFR). The solid line in both panels indicates the evolutionary track of simulation 1m14, of which the evolution of the slope is shown in Fig.~\ref{fig:slopemerger}. The mean slope and $\rho_{\rm h,SFR}$ of the progenitor galaxies (1dB) is indicated with a cross.
       }
\end{figure}
In order to consider the relation between the slope of the age distribution and the star formation rate density, we have used the same approach as in Sect.~\ref{sec:agedis} to determine a measure of the star formation rate density in galaxy mergers. For both galaxies, we determine the half-mass radius of the gas distribution and add the enclosed volumes, leaving out any overlap between both spheres. To avoid artificially low star formation rate densities, the tidal arms are omitted when calculating the half-mass radius by neglecting all material beyond 100~kpc from the centre of mass of the simulation. The plane of the fitted age distribution slope versus star formation rate density is shown in Fig.~\ref{fig:agesfhall} for all galaxy merger simulations, also including the data from the galaxy disc simulations (see Fig.~\ref{fig:dslope}). As explained above in the discussion of Fig.~\ref{fig:slopemerger}, the galaxy mergers do not follow the relation between slope and star formation rate density that holds for isolated disc galaxies. Instead, during starbursts they typically move to shallower slopes and higher star formation rate densities, i.e. up and to the right in Fig.~\ref{fig:agesfhall}. The large scatter on the points of the galaxy merger simulations arises because of the wide range of possible age distribution slopes over the course of a single merger, which is also present in Fig.~\ref{fig:slopemerger}. The scatter is also increased by our method of estimating a measure for the star formation rate density, which only allows for an order-of-magnitude analysis because it is sensitive to the global dynamical changes during the merger.

The typical evolution of a galaxy merger in the diagram of Fig.~\ref{fig:agesfhall} is illustrated by the evolutionary track of simulation 1m14, which goes through three phases. Initially, both galaxies reside on the relation for isolated disc galaxies (dotted line and cross). For simulation 1m14, this is not shown in Fig.~\ref{fig:agesfhall}, because it occurs too early on in the simulation and insufficient clusters exist in the fitted age range. The evolutionary track starts at the top middle of the diagram, during the first pericentre passage, when the star formation rate density is still intermediate ($\rho_{\rm h,SFR}\sim 10^{-4}~\msun~{\rm kpc}^{-3}$) and cluster migration and natural selection are important, resulting in a shallow age distribution. In between both pericentre passages, it returns to the relation for isolated discs because the discs evolve back to a quasi-isolated state as in Fig.~\ref{fig:dslope}, but with a slightly higher star formation rate density. This changes just before the final coalescence, when the density contrast between the starburst region and the surroundings becomes important again, moving the galaxy to the top right of Fig.~\ref{fig:agesfhall}.

\subsection{The cluster population of merger remnants} \label{sec:remnant}
\begin{figure}
\resizebox{\hsize}{!}{\includegraphics{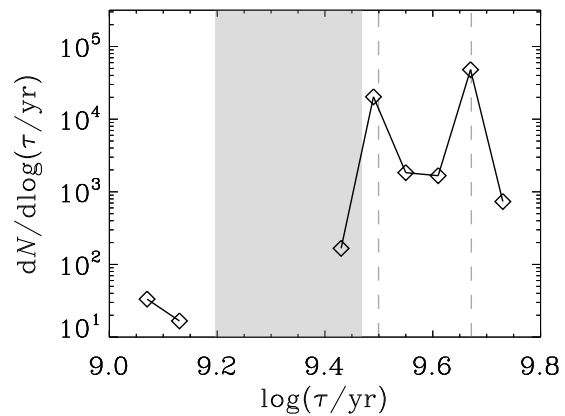}}
\caption[]{\label{fig:agedismerger}
Logarithmic age distribution ${\rm d}N/{\rm d}\log{(\tau/{\rm yr})}$ of the clusters with ages $\tau\geq 1$~Gyr in the merger remnant of simulation 1m14, shown for the snapshot at $t=4.9$~Gyr. The vertical dashed lines indicate the moments of first (right) and second (left) pericentre passage, while the shaded area marks the period over which the final coalescence occurs.
       }
\end{figure}
After a galaxy merger is completed and both galaxies have transformed into a single elliptical or S0 galaxy, the formation of stars and clusters ceases or proceeds at a low rate (${\rm SFR}<0.5~\msun~{\rm yr}^{-1}$). As a result, the vast majority of clusters in a merger remnant is old, with ages dating back to the first and second pericentre passages of the interaction. A first indication of when and where these clusters (the `survivors') were formed is obtained from their age distribution, which is shown in Fig.~\ref{fig:agedismerger} for the cluster population older than 1~Gyr of simulation 1m14. The age distribution shows that most of the survivors are formed approximately at the times of the first and second pericentre passages, just before the corresponding starbursts. During the last part of the coalescence, some more clusters are formed that survive the merger. Interestingly, no clusters with ages corresponding to the onset of the coalescence exist in the merger remnant, because the violent gas influx and the resulting high gas density shortens the lifetimes of the clusters that are formed under these conditions.

\begin{figure}
\resizebox{\hsize}{!}{\includegraphics{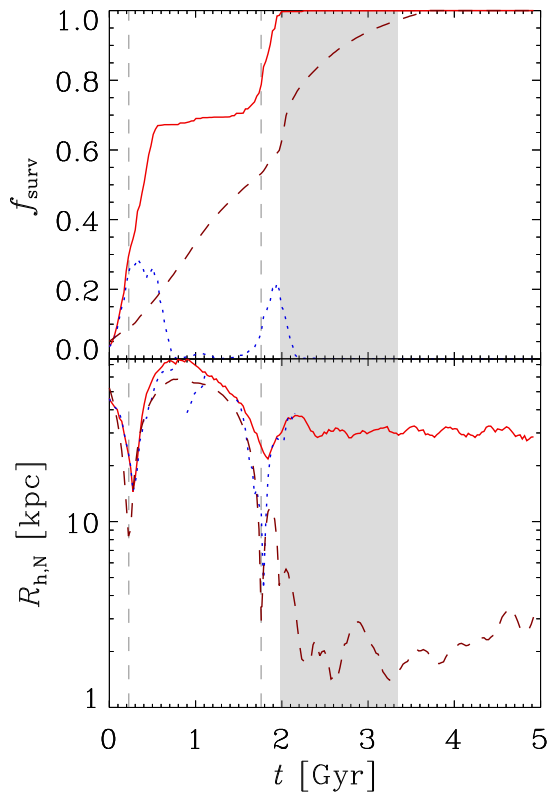}}
\caption[]{\label{fig:rgcevo}
(Cumulative) formation history and radial evolution of the clusters that will survive the galaxy merger of simulation 1m14, i.e. those that are still present at $t=4.9$~Gyr. {\it Top}: For each time $t$, the figure shows the fraction of the surviving cluster population that has been formed since the start of the simulation (red solid line) and the fraction that was formed during the 200~Myr preceding $t$ (blue dotted line). The dark red dashed line shows the cumulative fraction of star particles that have been formed since the start of the simulation. {\it Bottom}: Half-number radius of all present survivors (red solid line), of the survivors that were formed during the 200~Myr interval before time $t$ (blue dotted line), and of the star particles that have been formed since the start of the simulation (dark red dashed line). Stars and clusters formed in the range $t=4$--5 Gyr are ignored.
       }
\end{figure}
A more precise picture of the origin of the cluster population in the merger remnant is obtained by considering their cumulative formation history and the radial evolution of their population. This is shown in Fig.~\ref{fig:rgcevo}, which follows the time evolution of the (cumulative) relative formation history and the half-number radius for three groups of objects: all survivors formed since the start of the simulation (giving a cumulative fraction), the survivors formed during the last 200~Myr, and all star particles formed since the start of the simulation (also giving a cumulative fraction). Contrary to the half-mass radius of the gas in Sects.~\ref{sec:agedis} and~\ref{sec:agedismerger}, the half-number radius considered here is not defined with respect to the centre of the appropriate galaxy, but with respect to the centre of mass of the entire simulation.

The cumulative formation history of the survivors shows that each of both pericentre passages contributes about 30--60\% of the old cluster population in the merger remnant. The precise distribution of percentages depends on the orbital geometry of the merger and on the properties of the progenitor galaxies. In the case of simulation 1m14, which is shown in Fig.~\ref{fig:rgcevo}, the galaxies pass each other on near-polar orbits, yielding a weaker starburst than a head-on or co-planar encounter and leaving some gas for post-merger star formation. For more violent starbursts, all gas is consumed and no young clusters exist in the merger remnant.

The assembly history of the stellar mass is distributed over both pericentre passages in a way that is similar to that of the clusters, even though the first passage gives rise to a much smaller starburst than the second passage. The stellar mass formed in both passages is comparable because the duration of the first starburst exceeds that of the second. The most remarkable difference between the formation history of the star particles and the surviving clusters is that they are offset with respect to each other. The surviving clusters are typically formed at earlier times than the star particles, which was also mentioned in Sect.~\ref{sec:agedismerger} and the discussion of Fig.~\ref{fig:agedismerger}. Most of these survivors were ejected into the stellar halo during the pericentre passages and survived because halo clusters experience a lower disruption rate than clusters residing in the discs of both galaxies. These ejected clusters were formed before the starburst, because the onset of ejection into the halo precedes the moment of peak starburst intensity by $\sim 200$~Myr. The combination of an already enhanced star formation rate before the ejection and the increased survival chances of halo clusters implies that the ejected clusters constitute a large part of the post-merger population of survivors. 

The ejection of clusters can also be seen by considering the half-number radii of the system of (recently formed) surviving clusters and of the stars in Fig.~\ref{fig:rgcevo}. The pericentre passages of the two galaxies are visible as minima in the evolution of the half-number radius of the star particles. Already during the first passage, the half-number radius of the clusters exceeds that of the star particles, because the clusters that are ejected from the discs of both galaxies have higher survival chances than the clusters that stay confined to the discs. This effect becomes even more important during the second passage and final coalescence of the galaxies, during which the half-number radius of the clusters hardly changes, but the star particles end up in a much smaller volume. While this could suggest that {almost} no survivors are formed at small radii, the half-number radius of the recently formed surviving clusters proves the contrary. During and shortly ($\sim50$~Myr) after the second pericentre passage, the spatial distribution of the recently formed survivors (blue dotted line in Fig.~\ref{fig:rgcevo}) is as confined as the spatial distribution of star particles. This illustrates that the clusters may be formed in the galactic discs, but are subsequently ejected due to the dynamical interaction of the galaxies, increasing their chances for survival. At later times ($>50$~Myr after the pericentre passage), the survivors are formed at different locations than the stars, because the clusters that are formed in the centre of the starburst are disrupted. These two examples of natural selection imply that the spatial distribution of the star cluster population in merger remnants does not follow the distribution of the stars, but is spatially more extended.

\section{Discussion} \label{sec:disc}
In this section, we provide a summary and a discussion of the possible improvements and potential applications of our method.

\subsection{Summary} \label{sec:summary}
We have presented numerical simulations of isolated and merging disc galaxies, in which a sub-grid model for the formation and evolution of the entire star cluster population is included. The description for the star clusters is semi-analytic and includes a model for their internal dynamical evolution and the resulting changes of the stellar mass function within the clusters. The prescription for cluster disruption has been validated by comparing to $N$-body simulations of dissolving star clusters, giving good agreement. When considering individual clusters within our simulations, the tidal field strength and tidal shocks are found to have a clear effect on the mass loss histories of the clusters. This provides a verification of the presented method.

One of the key advantages of the model is that it shows how the disruption rate of clusters varies in time and space. We have used our disc galaxy simulations to assess the implications of this for characteristic properties of the cluster populations. We find that the mean age of the cluster population increases with galactocentric radius, because the disruption rate and the cluster formation rate are highest near the galactic centre. This is also found in observations of the clusters in M51 \citep{gieles05a} and the Milky Way \citep{vandenbergh80,froebrich10}. The relative contribution of tidal shocks to the disruption of star clusters is found to be $\sim 80$\%, which weakly increases with increasing gas fraction of the galactic disc. A similar value was found by \citet{lamers06a} from an analysis of clusters in the solar neighbourhood.

The combination of disruption due to two-body relaxation, tidal shocks, and their variation in time and space affects the slope of the cluster age distribution through two main mechanisms that lead to the same result. `Cluster migration', i.e. the motion of clusters away from their formation sites, and `natural selection', i.e. the higher survival probability of clusters in quiescent environments, both imply that the mean disruption rate decreases with age. In the extreme case, this can cause an age distribution with a single logarithmic slope of $-1$ over the majority of the age range, instead of the classical flat distribution at young ages combined with a steep decline at old ages. For isolated disc galaxies, the effects of cluster migration and natural selection are largest in low gas density galaxies, because these have higher gas density contrasts between star forming regions and their surroundings. Combining this with the relation between gas density and star formation rate density \citep{schmidt59,kennicutt98b}, we obtain a clear correlation between the star formation rate density and the slope of the disruptive (old) end of the age distribution, which is steeper for higher star formation rate densities.

Our simulations of galaxy mergers show that the disruption rates of clusters vary widely and depend on their orbital histories during the merger. The clusters that reside in the central regions of the galaxies are disrupted on short timescales, while clusters that are ejected into the stellar halo can survive for several gigayears. The mechanisms of cluster migration and natural selection are prevalent in galaxy mergers, because the environment of clusters strongly varies in time and space. As a result, the fitted slope of the cluster age distribution {(in the range $\log{(\tau/{\rm yr})}=7.7$--9) evolves from $-0.5$ or $-1$ during the starbursts, when the contrast between the concentrated star forming volume and its surroundings is largest, to $-2.5$ or $-3$} in between the pericentre passages, when the discs evolve back to a quasi-isolated state. This is a fundamental physical difference compared to isolated galaxies, in which the density contrast between star forming regions and their surroundings is largest for galaxies with low star formation rate densities.

The star clusters that survive the merger and populate the merger remnant are typically formed at the moments of the pericentre passages, i.e. slightly before the starbursts that occur during a galaxy merger. These clusters constitute a large fraction (30--60\% per pericentre passage) of the survivors for two reasons. Firstly, they are formed in large numbers, because the star formation rate already increases before the peak of the starburst. Secondly, during the pericentre passage, the formed clusters are ejected into the stellar halo, where the disruption rate is low and the survival chance is high. The clusters that are produced in the central region during the peak of the starburst are short-lived and disrupt before they can migrate to the halo. As a result, a peak in the star formation rate does not necessarily correspond to a peak in the cluster age distribution. Depending on the properties of the starburst and the time that elapsed since it occurred, both peaks will be offset with respect to each other.

This paper shows that the variability of the disruption rate in time and space has a pronounced impact on the properties of cluster populations in a range of galactic environments. It affects the spatial distribution of clusters, their age distribution, and the evolutionary histories of the clusters that survive until the present day. As discussed in Sect.~\ref{sec:intro}, it has been common practice in literature to adopt a single, ``mean'' disruption rate for the entire cluster population of a galaxy. While this approach holds many advantages due to its simplicity, we now see that the resulting cluster populations have very different properties than those ensuing from a more realistic setting, in which the effects of the formation, disruption, and orbital histories of the clusters are intertwined.

\subsection{Improvements} \label{sec:improve}
While the presented model gives a more detailed description of the formation and evolution of star cluster populations than before, there are several points at which it could be improved. We discuss five key improvements.

\begin{itemize}
\item[(1)] The current treatment for star formation uses one gas particle per spawned star particle. This implies that the particle mass limits the maximum cluster mass, because it is not possible to form clusters that are more massive than the star particle which they are part of (see Sect.~\ref{sec:clform}). As a result, it is not beneficial to increase the resolution of the simulation, because it will decrease the maximum cluster mass below the current value of $\sim 10^{5.8}~\msun$. Especially when considering galaxy mergers, in which clusters with masses around $10^7~\msun$ should be produced, improving this would be very relevant. We intend to include a group-finding algorithm in the near future, which will evaluate the Jeans criterion for groups of gas particles. This would enable the formation of a single star particle out of multiple gas particles, and will also allow us to increase the resolution of the simulations without compromising the mass range of the cluster population. In addition to giving a more realistic description of the star formation process, this would also enable us to resolve the ISM down to smaller scales, and improve the description of cluster disruption due to tidal shocks.
\item[(2)] Supermassive black holes (SMBHs) and the possible feedback from SMBHs are presently not included. The vast majority of star clusters resides in the range where the tidal field due to the SMBH can be neglected, so the disruption rate of star clusters is not directly affected by the omission of SMBHs. An indirect effect of the presence of SMBHs could be important in galaxy mergers, during which feedback from SMBHs may be responsible for the expulsion of all gas from the galaxy \citep{dimatteo05}. This would disrupt any gas discs that may reform in the merger remnant and would halt further formation of star and clusters. Because it is a second order effect for the problem we are addressing, and because there are currently no definitive models for SMBH feedback \citep{pelupessy07,sijacki10}, we have chosen to omit SMBHs in the present model. Whenever a more conclusive model for SMBH feedback becomes available, it will be included in our model.
\item[(3)] We have approximated the evolution of the half-mass radius of star clusters with a simple power law dependence on the cluster mass, fixing the normalisation and power law index by means of a comparison to $N$-body simulations of dissolving clusters on eccentric orbits. This is important, because the disruption timescale due to tidal shocks depends on the half-mass density. Using the adopted relation, we reproduce the disruption times found in the $N$-body simulations. Even though the relation is consistent with the theoretically expected relation in the `mass loss-dominated' regime from \citet{gieles10b}, a better approach would be to adopt a prescription for the half-mass radius that has a more extensive physical foundation. Unfortunately, current mass-radius relations in literature are based on the evolution of clusters in a smooth galactic potential, and depend on the galactocentric radius \citep{gieles10b}. While this is accurate for globular clusters on orbits with a low eccentricity, it does not work for clusters orbiting within a galactic disc or in galaxy mergers, where the tidal field is erratic due to the non-uniform distribution of the gas. An appropriate model for the evolution of the half-mass radius in such an environment could be obtained by feeding an erratic tidal field into $N$-body simulations of star clusters and monitoring their structural evolution. Such an analysis is well beyond the scope of the present work, and we will update the mass-radius relation whenever a better description becomes available. 
\item[(4)] At present, the model does not include a description for chemical enrichment, and consequently all clusters have the same metallicity. While this has a negligible effect on the mass evolution of the clusters, their photometry is affected (see \citealt{kruijssen08} for a quantitative analysis). {Moreover, including a prescription for the chemical evolution of the star cluster population would enable a better comparison with (spectroscopic) observations, in which chemical abundances can be established with a generally higher accuracy than other properties such as cluster ages. It would also allow us to investigate the relation between metallicity and other characteristics of the cluster population, and to improve the model for star formation, which depends on the chemical composition of the gas.} We aim to include a model for chemical enrichment in a future work.
\item[(5)] The cluster formation efficiency (CFE), i.e. the fraction of stars that is formed in a clustered form, is assumed to be constant. This implies that the exact value acts as a normalisation of the total number of clusters, leaving it as a free parameter. It is set to 90\% to obtain better statistics for the simulated cluster populations (see Sect.~\ref{sec:clform}). However, there have been suggestions that the CFE depends on the local environment, particularly on quantities like the star formation rate density \citep{goddard10}. The exact dependence of the CFE is still far from certain, but if there exists an environmental dependence, this would affect the cluster population by favouring the formation of clusters in certain parts of a galaxy. This could also have a secondary effect on the cluster population, because cluster disruption may also proceed differently in parts with an enhanced CFE. Again, an environmental dependence of the CFE will be included when it is better constrained, either from models or observations.
\end{itemize}

Apart from these main areas for improvement, we will keep updating the models as $N$-body simulations and observations of clusters in a broader range of environments become available.

\subsection{Applications} \label{sec:apply}
In order to trace the formation and evolution of galaxies using star cluster populations, it is necessary to investigate how different galactic environments affect the cluster population. Our model is a very suitable tool to gain more insight into this question, because it relates the evolution of each cluster to its (time-dependent) local environment. This implies a certain flexibility that allows us to apply the model to a broad range of galaxies. While a first analysis of the interplay between galaxies and their star cluster populations is already given in this paper, there are many more observables of the cluster population that should be investigated under different galactic conditions.

It would be particularly useful to understand the impact of galaxy mergers on cluster populations, because such an understanding enables the use of cluster populations to probe merger histories and the hierarchical assembly of galaxies. Mergers are recognized as important drivers of starbursts and corresponding cluster formation, which are fuelled by high gas densities. However, as is shown in this paper, a high gas density also implies a large disruption rate. It is not trivial to determine whether formation or destruction dominates. We have considered this question in \citet{kruijssen10} as a first application of the model, and find that the total number of clusters decreases during a merger, because the large gas densities result in more destruction than formation. Destruction is most prominent for the numerous clusters with low masses, whereas for the fewer massive clusters formation does dominate during certain episodes of the galaxy interaction. The corresponding change of the cluster mass function could be used as a tracer of the merger type.

By modeling specific, real galaxies, it is possible to explain observed properties of the cluster population and to predict its features that presently fall below the detection limit. Such case studies will also verify the model, and possibly provide constraints on aspects of the model that are currently uncertain (see Sect.~\ref{sec:improve}). For instance, by comparing the observed and modeled star formation rates and the number of clusters within a certain mass and age range, it will be possible to infer the cluster formation efficiency in a particular galaxy\footnote{Before being able to combine different observational studies to look at trends of the properties of the cluster population with galactic environment, the current dichotomy in literature between groups drawing different conclusions from the exact same observational data \citep[e.g.][]{chandar06,gieles07b} should be settled. No model can bridge such differences.}.

As is indicated in Sect.~\ref{sec:intro}, the disruption rate of star clusters is commonly assumed to be constant when deriving the star formation history (SFH) of a galaxy from its star cluster population. Although this approximation is convenient, the thus obtained SFH will differ from the actual one. The impact of the disruption time on the inferred SFH was recently illustrated by \citet[Fig.~4]{maschberger10}, who show that it depends on the adopted disruption rate to what extent the gap in the age distribution of clusters in the Large Magellanic Cloud is reflected in the inferred SFH. For their choice of disruption rates, the SFR in the age range corresponding to the age gap varies by about 1.5~dex, resulting in cases in which the SFR does and does not contain the age gap of the cluster age distribution. Because the conditions within an evolving galaxy vary widely, the impact of the time- and space-variation of the disruption rate are likely of the same order of magnitude. It is therefore essential to resolve how this variation may affect SFHs that are inferred from the star cluster population.

The formation and evolution of star cluster populations are the result of several mechanisms that act simultaneously, such as starbursts, feedback, tidal shocks, two-body relaxation, cluster migration, natural selection, and many other processes. While certain parts may still be uncertain, the current understanding of these mechanisms enables the modeling of the cluster population in a way that reflects the variability and complex nature of real galactic environments. Future applications of the model should therefore provide new clues to the (co-)evolution of galaxies and their star cluster populations.

\section*{Acknowledgments}
{We would like to thank the anonymous referee for inspiring and stimulating suggestions}. We are indebted to Oleg Gnedin for sending us his routines for tidal shock heating. Mark Gieles is thanked for interesting discussions and for providing some characteristic numbers for $N$-body King models. Holger Baumgardt kindly provided the evolution of the half-mass radius from his $N$-body simulations. We thank Nate Bastian, S$\o$ren Larsen and Esteban Silva Villa for helpful discussions. The calculations reported here were performed at the computing facilities of Leiden Observatory. This research is supported by the Netherlands Advanced School for Astronomy (NOVA), the Leids Kerkhoven-Bosscha Fonds (LKBF) and the Netherlands Organisation for ScientiÞc Research (NWO), grants 021.001.038, 639.073.803, and 643.200.503. JMDK acknowledges the kind hospitality of the Institute of Astronomy in Cambridge, where a large part of this work took place.

\label{lastpage}

\end{document}